\magnification=1200
\tolerance=5000
\null
\hyphenpenalty=2000
\def\degree{^{\circ}}
\def\SH{{\cal H}}
\nopagenumbers

\vskip0.5truein
\centerline{\bf Exoplanets or Dynamic Atmospheres?}
\centerline{\bf The Radial Velocity
and Line Shape Variations of 51 Pegasi and $\tau$ Bo\"otis}
\vskip0.7truein

\centerline{Timothy M. Brown, Rubina Kotak}
\centerline{High Altitude Observatory/National Center for Atmospheric Research
\footnote{$^*$}{The National Center for Atmospheric Research is Sponsored by
the National Science Foundation}}
\centerline{P.O Box 3000, Boulder, CO 80307}
\centerline{brown@hao.ucar.edu, rubina@astro.lu.se}
\vskip10pt

\centerline{Scott D. Horner}
\centerline{Department of Astronomy \& Astrophysics, 
Pennsylvania State University}
\centerline{University Park, PA 16802}
\centerline{horner@astro.psu.edu}
\centerline{and}
\vskip10pt

\centerline{Edward J. Kennelly, Sylvain Korzennik, P. Nisenson
\& Robert W. Noyes}
\centerline{Harvard-Smithsonian Center for Astrophysics, Cambridge MA 02138}
\centerline{tkennelly@cfa0.harvard.edu, sylvain@cfa0.harvard.edu,
nisenson@cfa0.harvard.edu,}
\centerline{noyes@cfa0.harvard.edu}

\vskip0.7truein
Received \_\_\_\_\_\_\_\_\_\_\_\_\_\_\_\_\_\_\_\_\_\_\_\_\_\_\_\_\_\_\_\_\_
\_\_\_\_\_\_\_\_\_\_\_\_\_\_\_\_\_\_\_\_\_\_\_\_\_\_\_\_
\vfill\eject

\centerline{\bf Abstract}

The stars 51 Pegasi and $\tau$ Bo\"otis show radial velocity variations
that have been interpreted as resulting from 
companions with roughly jovian mass and orbital periods of a few days.
Gray (1997) and Gray \& Hatzes (1997) reported
that the radial velocity signal of 51 Peg is synchronous with
variations in the shape of  the line $\lambda$6253 Fe I;
thus, they argue that the velocity signal arises not from a companion
of planetary mass, but from dynamic processes
in the atmosphere of the star, possibly nonradial pulsations.
Here we seek confirming evidence for line shape or strength variations
in both 51 Peg and $\tau$ Boo, using $R=50000$ observations taken
with the Advanced Fiber Optic Echelle.

Because of our relatively low spectral resolution,
we compare our observations with Gray's line bisector data by fitting
observed line profiles to an expansion in terms of orthogonal
(Hermite) functions.
To obtain an accurate comparison, we model the emergent line profiles
from rotating and pulsating stars, taking the instrumental point spread
function into account.
We describe this modeling process in detail.

We find no evidence for line profile or strength variations
at the radial velocity period
in either 51 Peg or in $\tau$ Boo.
For 51 Peg, our upper limit for line shape variations with
4.23-day periodicity is small enough to exclude 
with 10$\sigma$ confidence the bisector curvature
signal reported by Gray \& Hatzes;
the bisector span and relative line depth signals reported by
Gray (1997) are also not seen, but in this case
with marginal (2$\sigma$) confidence.
We cannot, however, exclude pulsations as the source of 51 Peg's
radial velocity variation, because our models imply that 
line shape variations associated with pulsations should be much
smaller than those computed by Gray \& Hatzes;
these smaller signals are below the detection limits 
both for Gray \& Hatzes' data and for our own.

$\tau$ Boo's large radial velocity amplitude and $v \sin i$ 
make it easier to test for pulsations in this star.
Again we find no evidence for periodic line-shape changes,
at a level that rules out pulsations as the source of the
radial velocity variability.
We conclude that the planet hypothesis remains the most likely
explanation for the existing data.
\vfill\eject
\pageno=2
\footline={\hss\tenrm\folio\hss}

\parindent=0pt
{\bf I. Introduction}
\parindent=20pt

Since the reported detection of a low-mass companion to the Sun-like
star 51 Peg (Mayor \& Queloz 1995), a substantial and growing number
of stars have been shown to display similar radial velocity (RV) variations.
If the observed RV signals are interpreted as resulting from
the gravitational attraction of an orbiting body, then all of the
companion so far detected have masses between roughly 0.5 and 10
jovian masses.
Most of the stars in question 
(51 Peg, Mayor \& Queloz 1995; $\rho^1$ Cnc, $\tau$ Boo, $\upsilon$ And,
Butler {\it et al.} 1997;
$\rho$ CrB, Noyes {\it et al.} 1997)
show sinusoidal RV signals with
periods of 40 days or less.
In this paper we consider 2 of these stars:
51 Peg (G2-3 V), with a RV period and amplitude of 4.231 d and 56 m s$^{-1}$,
and $\tau$ Boo (F7 V), 
with a period and amplitude of 3.313 d and 468 m s$^{-1}$.
For Sun-like stars these properties imply circular orbits with
semimajor axes smaller than 0.25 astronomical units (AU).
Other stars (HD114762, Latham {\it et al.} 1989;
47 UMa, 70 Vir, Butler \& Marcy (1996), Marcy \& Butler (1996); 
16 Cyg B, Cochran {\it et al.} 1997)
show RV variations that are non-sinusoidal or that have 
longer periods, implying elliptical and/or larger orbits.

The appearance of ``hot Jupiters'', objects with roughly jovian mass
in orbits smaller than that of Mercury, is a puzzle and a challenge
to theories of solar system formation (Boss 1995).
Recently Gray (1997) (henceforth G97) and Gray and Hatzes (1997)
(henceforth GH)
have proposed a straightforward solution to this
problem, namely that the planet orbiting 51 Peg
does not exist.
By implication, other exoplanets with similar properties are also suspect.
G97 reports evidence for changes in the shape of the
spectrum line $\lambda$6253 Fe I and in the relative strength of the low- and
high-excitation potential lines $\lambda$6253 Fe I and $\lambda$6252 V I,
both changes being synchronous with the RV signal.
GH describe similar variations in the curvature of the $\lambda$6253
line bisector.
Such changes  are definitely not expected as a result of orbital
accelerations;  Gray interprets them as evidence for dynamical processes
in the atmosphere of 51 Peg
(perhaps nonradial pulsations, henceforth abbreviated NRP).
If the line shape or strength variations reported by G97 and GH are
indeed present, then it is likely that they are also the cause
of the RV variations, and one can no longer make
a case for the existence of most of the current crop of exoplanets.

In what follows we aim to address two questions connected with
the contention in G97 and GH,
using observations of 51 Peg and $\tau$ Boo.
First, can we use existing data to confirm Gray's observations
of 51 Peg?
Second, can evidence for similar line-symmetry variations be found
in other stars reported to harbor hot Jupiters?
In particular, we may learn something by examining $\tau$ Boo,
which has 9 times the RV variability of 51 Peg,
and which also rotates faster ($v \sin i = 15$ km s$^{-1}$
for $\tau$ Boo {\it vs.} less than 3 km s$^{-1}$ for 51 Peg).
In order to address this second question, it is necessary 
to calculate the observable consequences of NRPs on the spectrum lines
of stars whose line profiles suffer significant rotational broadening.
We shall be concerned with both the mean Doppler shift suffered
by spectrum lines and with changes in line shapes (particularly
asymmetries) induced by pulsations.

A summary of our
Doppler shift and line shape results for 51 Peg and
in $\tau$ Boo has been provided elsewhere (Brown et al. 1998);
in the present paper
we shall therefore concentrate upon the details of the analysis
of these data, which were necessarily slighted in the previous work.
We find important differences between our results and those of G97
and GH, with respect both to the observed behavior of 51 Peg and
also to the behavior of kinematic models of the emergent line profiles,
should the star be pulsating.
In what follows we therefore describe our procedures in
detail, so that their nature and justification
may be as clear as possible.
The remainder of the paper is laid out as follows:
Section II  describes the physical basis upon which line-shape
variations arising from NRPs may be estimated, and identifies
some of the likely limitations to our ability to make these estimates.
In section III we discuss our strategy for using relatively low-resolution
(${ \lambda / \delta\lambda} = R = \ 50000$) spectra
to learn something about the shape and symmetry variations of
stellar absorption lines.
We give considerable attention to the expansion of the line profiles
in terms of Hermite functions, and to the relation between this
expansion and the more familiar line bisector analysis.
Section IV is a description of the observations of 51 Peg and $\tau$
Boo.
Section V describes the analysis (and particularly the time-series analysis)
of the derived parameters of line shape and strength.
We find here no positive evidence for the presence of NRPs on these
two stars, and we are able to place interesting limits on the possible
amplitudes of such pulsations.

\vskip12pt
\parindent=0pt
{\bf II. Physical Background and Notation}
\parindent=20pt

Gravitational attraction by an orbiting companion is an attractive
hypothesis to explain the radial velocities of stars such as 51 Peg,
because the effects are exactly calculable using a model with only
a few free parameters.
Radial velocity variations that are sinusoidal in time may be explained
in terms of circular orbits, in which case the only parameters are
the masses of the two orbiting bodies, 
the semimajor axis of the orbit,
and the inclination of the orbital axis to the line of sight.

The observable effects of NRPs are more complex to describe,
depend upon more parameters, and hence are prone to a more ambiguous
interpretation than are those of orbiting bodies.
This is because stellar pulsations manifest themselves as time-varying
perturbations in the flow and thermodynamic fields at the stellar
surface, with complicated spatial structures and with relationships
between physical variables that depend upon the physics of the oscillation
mode in question.
Uncertainties also arise because of the unknown orientation of the
modal pattern with respect to the line of sight, because of the
poorly-known variation of mode amplitude with depth in the stellar
atmosphere, and because of possible errors in the effects of stellar
limb darkening on the oscillating part of the emergent line profiles.
In this section we shall attempt to give some idea of the range of
possible physical processes, but we shall treat in detail only the
simplest cases, namely pressure- and gravity- (p- and g- ) modes
in stars where nonspherical components of the structure are
negligibly small.
Assuming near spherical symmetry implies that the stellar rotation period 
should be long compared to the oscillation period of interest, and that
magnetic fields should be unimportant everywhere.
These conditions are rather restrictive;
for instance, $\tau$ Boo does not satisfy the slow rotation condition.
Even imposing these conditions, however, it is surprisingly difficult
to make statements of a general nature about the observed line profiles.
Unavoidably, many of our conclusions will therefore be statistical.
The reader should keep in mind these uncertainties, to say nothing of
the more serious ones that may arise from a complete misidentification
of the relevant physics.

Pulsations of spherical stars have been an active subject of study
for many years; for descriptions of progress in this field
see the review articles by Kurtz (1990), Brown \& Gilliland (1994),
Gautschy \& Saio (1995), and references therein.
Different pulsation modes are distinguished by 
the restoring force that dominates
in returning displaced fluid elements to their equilibrium positions,
and by the mode's 3-dimensional spatial structure.
Because of the spherical symmetry of the equilibrium state,
the angular part of the eigenfunctions of any p- or g-mode 
can be described in terms of 
a single spherical harmonic 
$$
Y_{\ell}^m (\theta, \phi ) \ = \ P_{\ell}^m (\cos \theta ) e^{im\phi } \ ,
\eqno (1)
$$
where $\theta$ is the colatitude, $\phi$ is the longitude,
$P_{\ell}^m$ is an associated Legendre polynomial
normalized so that its maximum value is unity,
$\ell$ is the $angular$ $degree$ of the mode, and $m$ is the
mode's $azimuthal$ $order$, {\it i.e.,} the projection of $\ell$ onto
a chosen axis.
For a star with true spherical symmetry the axis may be chosen arbitrarily;
for slowly rotating stars it is convenient to align the axis of the
spherical harmonic coordinate system with the rotation axis.
The vertical ($r$) and angular ($\theta$, $\phi$)
components of the oscillating velocity field are then given by
$$
V_r \ = \ V_p Y_{\ell}^m \ ,
\eqno (2)
$$
$$
V_{\theta} \ = \ V_p k {d \over {d \theta}} Y_{\ell}^m \ ,
\eqno (3)
$$
$$
V_{\phi} \ = \ V_p {k \over {\sin \theta}} {d \over {d \phi }} Y_{\ell }^m \ ,
\eqno (4)
$$
where we term $V_p$ the $pulsation$ $amplitude$ and $k$ is 
related to the ratio of
horizontal to vertical amplitudes.

To identify an oscillation in a 3-dimensional domain,
one requires 3 indices.
Two of these are provided by $\ell$ and $m$;
the third is usually denoted $n$, and counts the number of nodes
in the mode eigenfunction along the stellar radius.
Since $n$ relates to the depth dependence of the mode,
and stellar spectroscopic observations usually respond to a very
limited range of depths within the stellar photosphere,
$n$ is ordinarily not accessible to direct observation.
Rather, $n$ is usually inferred from the oscillation frequencies,
which always depend upon it.

The magnitude of the angular derivatives in Eqs. (3-4)
grows in proportion to $\ell$.
Therefore, $V_p$ is a good indicator of the speed of surface flows
only if the product $k \ell$ is smaller than unity.
If $k \ell$ is large then horizontal motions dominate vertical ones,
and the surface speeds scale with $\ell$ when $k$ and $V_p$ are
held fixed.
The radial velocity periods that concern
us are all longer than 75 hours, which is much longer than the
(roughly 1 hour) characteristic dynamical time for Sun-like stars.
Thus, as already noted by GH,
one expects that pulsations with the observed periods
should have $k \geq 10^4$, and NRPs
should be characterized by flows that are almost entirely horizontal.

Pulsations cause fluctuations not only in surface velocity fields,
but also in temperature and hence brightness.
Photometric observations of 51 Peg and $\tau$ Boo show that their
brightness variations at the RV period are weak or absent
(Baliunas {\it et al.} 1996).
But because of the long periods involved,
these observations do not imply a strong constraint
on the nature of possible oscillations.
For low-amplitude NRPs (for which the radiating surface area of the
star remains sensibly constant),
the most important source of brightness variations
is usually the alternate compression and rarefaction of the gas
during the oscillation cycle.
The fractional compression during an oscillation cycle is
approximately the ratio of the displacement to the density scale height.
If the motions were adiabatic, the temperature variation
would be of the same order.
But the radiative cooling times are only a few seconds in the photospheres
of cool stars, so that the actual behavior is far from adiabatic.
In the Sun, for instance, where the scale height is 150 km,
a pulsation with 300 s period and
a displacement amplitude of 7.5 m  causes a fractional compression
of $5 \times 10^{-5}$.
Because of the short radiative cooling time, however,
the resulting relative temperature
variation is only about $7 \times 10^{-7}$,
with a corresponding relative intensity change of $3 \times 10^{-6}$
(Toutain \& Fr\"olich 1992).
At much longer periods, the surface temperature signal should be
correspondingly reduced.
The solar supergranulation is an example of such a long-lived flow.
Supergranules exhibit horizontal velocities of about 500 m s$^{-1}$,
lifetimes of a day or so, and no measurable temperature variation,
apart from incidental hot spots associated
with the magnetic structures that they sweep up.

The modes most commonly observed in stars are p-modes, in which the
restoring force is pressure.
These are essentially sound waves,
so an upper limit to p-mode periods is given by the sound travel
time across the star.
For cool main-sequence stars, this time is about
an hour.
Since the observed radial velocity periods are all 75 hours or more,
p-modes cannot explain them, and need not concern us further.

In gravity (or g-) modes, the restoring force is buoyancy, which necessarily
involves horizontal gradients in density.
For this reason, spherically symmetric ($\ell = 0$) g-modes do not exist,
but nonradial modes (those with $\ell \geq 1$) are possible.
Gravity modes are less commonly seen in stars than are p-modes,
but they do exist, especially in stars where gravity is strong and
the vertical density stratification is pronounced.
The propagation of g-modes depends on the presence of a buoyant
restoring force,
which is absent or negative in convectively unstable regions,
so that gravity waves cannot propagate within stellar convection zones.
Stars like the Sun, with moderately deep surface convection zones,
are unlikely to show surface evidence of g-modes with large
values of $\ell$.
To date, g-modes have not been detected in the Sun, down to an amplitude
limit of less than about 1 mm s$^{-1}$ in the horizontal velocity 
at periods of a few hours (Fr\"olich \& Andersen, 1995).

If g-modes could be seen in the Sun, their periods would be determined
by the density stratification of the solar interior, and by their
spatial structure.
Asymptotically ({\it i.e.,} for $n \ \gg \ \ell$), 
g-mode periods $T (n, \ell )$ obey
$$
T(n, \ell ) \ = \ {{T_0 (n \ + \ \ell /2 \ + \ \delta )} \over L} \ ,
\eqno (5)
$$
where $\delta$ is a constant of order unity, $L = \ell (\ell + 1)$,
and $T_0$ is the asymptotic period, which depends upon an integral
of $N$ throughout the interior of the star (Tassoul 1980).
For the Sun, $T_0$ is about 1 hour;
it tends to decrease somewhat with stellar age, as the compositional
stratification of the core increases.
This value of $T_0$ implies that 
a g-mode with a period of 100 hours in a Sun-like star
would have a radial order $n$ of approximately 100.
How to excite such a mode without at the same time exciting its
neighbors with $n=99$ and $n=101$ would be a puzzle.

Finally, we consider the possibility that the observed cyclic Doppler variations
result not from planets, nor from NRPs,
but rather from Something Else.
G97 argues that the presence of line shape changes in 51 Peg
excludes the planet hypothesis, even if a viable alternate
hypothesis should be unavailable.
This general line of argument is undoubtedly correct.
On the other hand, there is an understandable reluctance to accept
explanations in the Something Else category, because they are so hard to test.
One must therefore take care before invoking an unknown process
to ensure that the evidence really requires it.
For this reason, we shall concern ourselves mostly with
comparisons between closely related quantities observed by Gray
and by ourselves, and with
the feasibility of interpretations in terms of known sorts of NRPs.
In the Discussion section below, we shall comment briefly on 
a marginally well-defined hypothesis, namely that
tides raised by an orbiting companion might cause asymmetries in
spectrum lines.

\vskip20pt
\parindent=0pt
{\bf III. Analysis Strategy and Line Profile Simulation}
\parindent=20pt

G97 and GH
presented two kinds of evidence to show that intrinsic
variations with a 4.23-day period are occurring in 51 Peg.
Both kinds are based upon a set of 39 low-noise, high-resolution
($R \ge 10^5$) spectra of 51 Peg obtained between 1989 and 1996.
First, G97 measured the shape of the line bisector of 
$\lambda$6253 Fe I, characterizing each bisector by its $span$, defined
to be the displacement (in m s$^{-1}$) between the bisector
position  at fractional flux levels (relative to the continuum flux)
of .50 and .90.
When phased with 51 Peg's radial velocity period (taken to be 4.2293 days),
these span values are fit by a sinusoid whose amplitude is
roughly 40 m s$^{-1}$
and whose phase (modulo a 180$\degree$ ambiguity
in the meaning of the bisector span phase) is the same as that of
the RV signal.
G97's observations in 1996 are better distributed than in previous
years, and show a larger variation of the bisector span 
than does the entire data set.
GH revised the analysis of G97 to concentrate on the line bisector
$curvature$, defined as the difference between the bisector span
measured between $F/F_c = $0.85 and 0.71, and that between
$F/F_c =$ 0.71 and 0.48.

Gray's second line of evidence consisted of a similar time series of the
ratio of the central depth of the line Fe I $\lambda$6253 to that of V I 
$\lambda$6252.
By a phased superposition method identical to that employed for the
bisector span, he found that this ratio varied sinusoidally around a
mean value of about 21.6\%, with an amplitude of 0.6 \%
and a phase that leads the bisector span phase by about 50$\degree$.
As noted by Gray {\it et al.} (1992),
the lower levels of these two lines have quite different excitation
potentials, so this line strength ratio is an indicator of
effective temperature.
The sensitivity estimated by Gray {\it et al.} (1992) is about 10 K per 1\%
variation in the depth ratio.
If interpreted as a temperature effect, the line depth ratio would
therefore suggest a temperature variation of about $\pm$ 6 K,
or $\delta T/T \simeq 10^{-3}$.
G97 explicitly warned against interpreting the line depth
ratio in terms of temperature alone, but offered no alternatives.

\vskip12pt
\parindent=0pt
{\it A. Line Profile Description in Terms of Hermite Functions}
\parindent=20pt

We wish to test these results using the
extensive set of $R = 50000$ spectra of 51 Peg taken with the Advanced
Fiber Optic Echelle (AFOE; Brown {\it et al.} 1994) since
1995.
It is straightforward to do so for the line depth ratio, because this
quantity is relatively insensitive to the wavelength resolution
employed.
Line bisector measurements are, however, notoriously sensitive to the
resolution ({\it e.g.,} Dravins, Lindgren \& Nordlund 1981).
Comparisons between line bisectors measured with $R = 50000$ and
with $R = 100000$ are therefore of little value,
and a more robust (though correspondingly less sensitive) measure of
line shape and symmetry must be used.
To this end we have chosen to expand the observed line profiles in
terms of suitable orthogonal functions.
Such an expansion is similar in concept to the ``moment method''
for describing the behavior of line-profile variables ({\it e.g.}
Balona 1986).
This approach has three advantages:
it allows comparisons between observations
taken with different resolution,
it permits a simple analysis of the average shape changes
occurring in an ensemble of many spectrum lines,
and it encourages an analytic approach to the calculation of line
bisectors from a prescribed distribution of radial velocity on the
surface of a star.
The last capability is useful both for general insight into the problem
of interpreting line shapes, and for performing elementary checks
of our numerical calculations of the line profile shapes.

Observed line profiles are fairly well approximated by Gaussians.
Thus, it is natural to express them in terms of Hermite functions $\SH_n 
({x / \sigma})$,
defined by
$$
\SH_n({x \over \sigma}) \ = 
\ {N_n \over \sqrt \sigma} \ exp(-{x^2 \over 2 \sigma ^2}) \ 
H_n({x \over \sigma}) \ ,
\eqno (6)
$$
where $H_n ({x / \sigma})$ is a Hermite polynomial as defined by
Abramowitz and Stegun (1972),
$\sigma$ is the usual measure of the line width,
and $N_n$ is a normalization factor given
by $N_n = ( 2^n n! \sqrt \pi )^{-1/2}$.
Normalized in this way, the functions $\SH_n$ satisfy the usual
orthogonality relation:
$$
\int \limits _{- \infty} ^\infty \SH_k \SH_j dx \ = \ \delta_{jk} \ .
\eqno (7)
$$

To fit an isolated absorption line profile, 
we represent the intensity $I(\lambda )$ in the neighborhood of a
fiducial wavelength $\lambda_f$ by
$$
I(\lambda) \ = \ C[1 + S (\lambda - \lambda_f)] 
\left [ 1 \ - \ D \left ( \SH_0 ({\lambda -
 \lambda_c \over \sigma}) \ + \ \sum \limits _{i=3} ^n h_i \SH_i ({
 \lambda - \lambda_c \over \sigma}) \right ) \right ] \ .
 \eqno (8)
$$
The parameters in this expansion are {\it C, S, D,} $\sigma$, $\lambda_c$, and 
the $n-2$ coefficients $h_i$.
Parameters $C$ and $S$ give the mean intensity and slope of the continuum
in the neighborhood of $\lambda_f$.
The fractional line depth (lying between 0 and 1) is $D$, and the line center
position is $\lambda_c$
The coefficients $h_n$ describe the departures of the line shape from
a simple Gaussian.
Because they are multiplied by $D$, 
and themselves multiply functions of $\lambda - \lambda_c$,
these coefficients are line shape
parameters that are independent of the line depth and position.
Notice that $h_1$ and $h_2$ are implicitly set to zero.
The function $\SH_1$ is proportional to ${d\SH_0 / d \lambda}$,
so the contribution of $\SH_1$ may always be absorbed in $\lambda_c$.
Similarly, the contribution of $\SH_2$ is absorbed in $\sigma$,
the lowest-order linewidth parameter.
Higher terms in the series cannot be represented in these ways,
and must be dealt with explicitly.

For fitting to observed spectra where there may be many absorption lines,
we generalize Eq. (8) to allow the $j^{th}$ spectrum line to be represented
by its own depth $D_j$ and its own center position $\lambda_{cj}$.
All lines share the same shape parameters $\sigma$ and 
$h_i$, however, and all are
measured relative to their local continuum, 
which is still represented as a linear
function of wavelength.
The resulting multi-line model is described by
$$
I(\lambda) \ = \ C[1 \ + \ S(\lambda - \lambda_f)] 
\left [ 1 \ - \ \sum \limits _j
D_j \left ( \SH_0( {\lambda - \lambda_{cj} \over \sigma} ) \ + \ 
\sum \limits _{i=3} ^n h_i \SH_i ( {\lambda - \lambda_{cj} \over \sigma } )
\right ) \right ] \ .
\eqno (9)
$$
The approximation that all lines have the same shape parameters is
valid so long as depth-independent kinematic influences 
on the line shape dominate
those of atomic and radiative-transfer effects ({\it e.g}, thermal
and Stark broadening, or depth-dependent correlations between velocity
and intensity).
In practice, this means that one must avoid using wavelength ranges
that contain very strong lines.

We now consider how the Hermite coefficients $h_i$, describing line shapes,
relate to the more
familiar parameterization in terms of line bisectors.
Suppose one starts with a Gaussian line profile $\SH_0(\lambda, \sigma )$
and distorts it by the addition of a small admixture of higher-order
Hermite functions.
For the present purpose it is convenient to hold $\lambda_c$
and $\sigma$ fixed, so that a complete description of the perturbed
profile requires terms involving $\SH_1$ and $\SH_2$.
We define the {\it relative line intensity} $U$ to be the (positive)
difference between the continuum and the line intensity,
normalized by the magnitude of $D$.
It is given by
$$
U( \lambda) \ = \  ( \sigma {\sqrt \pi})^{1/2} \left ( \SH_0 (\lambda , \sigma )
\ + \ 
\sum \limits _{i=1} ^n h_i \SH_i (\lambda , \sigma) \right ) \ \ ,
\eqno (10)
$$
where by assumption all $h_i \ll 1$.
What is the effect of nonzero coefficients $h_i$ on the line bisector?
From Fig. 1, it is evident that to first order the displacement 
$\delta B$ of the bisector at a given value of $U$ is
$$
\delta B (U) \ = \ -{1 \over 2} \left ( {\delta U_- \over dU/d\lambda_-} \ + \  
{\delta U_+ \over dU/d\lambda_+} \right ) \ ,
\eqno (11)
$$
where $\delta U_-$ ($\delta U_+$)
is the depth perturbation applied at the blue (red) side of the
profile at wavelengths corresponding to $U$ in the unperturbed profile, and 
$dU/d\lambda_{\pm}$ is the wavelength derivative of the intensity at
those points.
We approximate the derivative by that of the unperturbed profile,
which may be written (using the definitions of $\SH_0$ and $\SH_1$, and
dropping the explicit dependence of $\SH_n$ on $\sigma$):
$$
dU/d\lambda \ = \ - ( \sigma {\sqrt \pi})^{1/2} {\lambda \over \sigma^2}
\SH_0 (\lambda) \ = -{\sqrt {\pi \over 2 \sigma}} \SH_1 (\lambda) \ \ ,
\eqno (12)
$$
since $d\SH_0/d\lambda = (\sqrt 2 \sigma)^{-1} \SH_1$.
This gives
$$
\delta B (U) \ = \ {\sigma { \sqrt 2 \sum \limits _{i=1} 
^n h_i [ \SH_i (\lambda_-) \ - \ 
\SH_i (\lambda_+) ] } \over \SH_1 (\lambda_-)} \ \ ,
\eqno (13)
$$
where $\lambda_{\pm}$ are the two wavelengths at which the unperturbed
line profile has relative line intensity equal to $U$.

The functions $\SH_i$ have definite parity about the center of the line:
those of even $i$ are symmetric, and of odd $i$ are antisymmetric.
Thus, only the odd $i$ contribute to $\delta B$;
those with even $i$ modify the shape of the line but not its symmetry,
leaving the bisector position unchanged.
Moreover, a perturbation equal to $h_1\SH_1$ yields $\delta B \ = 2 \sqrt 2
\sigma
h_1 \ = \ {\rm constant}$, in effect translating the line profile without
changing its shape.
The size of the displacement is proportional to the 
coefficient $h_1$, and proportional to the line width parameter $\sigma$.
The lowest-order perturbation that can change the shape of the bisector
is then $\SH_3$.
If the perturbation is entirely of this form, then by writing out
$\SH_3$, one obtains
$$
\delta_3 B (U) \ = \ {\sigma \over \sqrt 3} \ h_3 \  
\left [ 2 \left ( {\lambda \over \sigma} \right )^2 - 3 \right ] \ .
\eqno (14)
$$
Since the unperturbed $U = \exp (- \lambda^2/2\sigma^2)$, 
Eq. (14) may be written
explicitly in terms of the relative line depth $U$:
$$
\delta_3 B (U) \ = \ - {\sigma \over \sqrt 3} \ h_3 \ 
( 4 \ln U \ - \ 3 ) \ \ .
\eqno (15)
$$

Shortly we shall show that for pulsating stars in which
the sum of the oscillating velocities and $v \sin i$ is small compared
to the $\sigma$ of the intrinsic line profile,
Eq. (10) provides a rapidly-converging description of the true line profile.
In such cases, we therefore expect pulsations to cause bisector variations
that are predominantly displacements that do not change the bisector shape
combined with those that are a logarithmic function
of $U$.
The relative magnitude of these two components is ill-defined, because
both $\SH_1$ and $\SH_3$ contribute to the displacement.
But as long as $\SH_3$ dominates the $U$-dependent part, one can predict
how measures of the bisector shape should behave.
If $v \sin i$ is larger than $\sigma$ (as in $\tau$ Boo),
then higher terms in the Hermite series may be important.
In what follows, we shall therefore retain the terms $h_4$ and $h_5$,
even though they are not necessary to describe 51 Peg's spectrum.

In Fig. 2 we illustrate the $span$ and $curvature$ parameters of bisector
shape that we shall use throughout this paper.
These are defined in velocity units, and the relative line intensities 
that we sample $\{d_1, d_2, d_3\} = \{.87, .48, .25\}$ 
are chosen to be the same as those
used by Hatzes (1997) and GH,
assuming a central line flux of 0.6 (measured in units of the continuum
flux).
We denote the displacement of the line bisector at relative line
intensity $d$ as $B(d)$.
The span $S_b$ and curvature $C_b$ are then expressed as
$$
S_b \ = \ B(d_1) - B(d_3) \ \ , \eqno (16)
$$
$$
C_b \ = \ 0.63 B(d_1) + 0.37 B(d_3) - B(d_2) \ \ . \eqno (17)
$$
The curvature is therefore the difference between $B(d_2)$ and the
linear interpolation between $B(d_1)$ and $B(d_3)$.
Given these definitions, we may expect
$$
S_b \ = \ {4 \over \sqrt 3} \ \sigma \ h_3 \ \ln \left ( 
{.87 \over .25} \right ) \ \simeq \ 
2.88 \ \sigma \ h_3 \ \ ,
\eqno (18)
$$
and
$$
{C_b \over S_b} \ = \ {{- \ln (.48) \ + \ .37
\ln (.87) \ + \ .63 \ln (.25) } \over { \ln (.85) \ - \ \ln (.25) }}
\ \simeq \ -0.16 \ \ .
\eqno (19)
$$

The foregoing paragraph assumed rapid convergence of the expansion
(Eq. 9) for the line profile shape, so that terms in the series
higher than $\SH_3$ could be ignored.
Is this assumption justified for 51 Peg?
To address this question, we must consider how to model the line
profiles produced by stars that both rotate and pulsate.
This modeling problem has been discussed in detail by many authors
({\it e.g.,} Vogt \& Penrod 1983, Kambe \& Osaki 1988,
Hatzes 1996, Schrijvers {\it et al.} 1997); 
here we repeat only the essentials of the 
formulation of the problem.

The line profile seen by a distant observer is the superposition of
the profiles emitted by each infinitesimal element of the star's
surface, with the contribution from each element Doppler-shifted
according to the line-of-sight component of its velocity,
and weighted according to the line strength appropriate to its
position on the stellar disk.
In a truly complete description of the process, one would 
use an appropriate stellar atmospheres model to take
account of the variation of line shape with position on the
stellar disk.
In what follows we shall not adopt this refinement,
assuming rather that the spectrum line is adequately
described by a profile whose strength and
center wavelength, but not shape, are functions of disk position.
Thus, the emergent line profile $I_o(\lambda)$ is described by
$$
I_o(\lambda ) \ = \ \int \ dx \ dy \ Q(x,y) \  
G(\lambda [1 \ + \ {v(x,y) \over c}]) \ \ ,
\eqno (20)
$$
where the integral is taken over the area of the stellar disk,
$Q(x,y)$ describes the spatial variation of the line strength,
$G(\lambda)$ is the line profile produced by stellar material
at rest with respect to the observer,
$v(x,y)$ is the line-of-sight component of the velocity,
and $c$ is the speed of light.

For the numerical simulations discussed below, we evaluated
the integral in Eq. (20) by numerical integration on a suitably
dense grid of points $(x,y)$.
It is instructive, however, to recast Eq. (20) by lumping
together all of the area elements on the stellar disk having
similar line-of-sight velocity $v$, and then integrating over
$v$.
If we ignore the change in apparent line width with
varying Doppler shift (a good approximation so long as $v/c$ is
small),
then the resulting integral may be written as
$$
I_o(\lambda ) \ = \ \int \ dv \ W(v) G(\lambda + v{\lambda_0 \over c}) \ \equiv
\ {c \over \lambda_0} \int d\xi \ W(\xi) G(\lambda_0 + \xi) \ \ ,
\eqno (21)
$$
where $\lambda_0$ is the central wavelength of the unshifted line,
$\xi = \lambda_0 v/c$ is the displacement from line center transformed
from velocity to wavelength units,
and $W(\xi)$ is the Doppler-shift density, representing the intensity-weighted
area of the star with line-of-sight velocity lying between $\xi$ and $\xi+d\xi$.
If we assume $G(\lambda_0 + \xi)$ to be symmetric in $\xi$ (which
is usually the case), 
then $I_o(\lambda)$ is seen to be the convolution of
the unshifted line profile with the Doppler-shift density $W$.

Now we wish to express $I_0(\lambda)$ in terms of the Hermite functions
$\SH_i(\lambda,\sigma)$.
To do this, we expand $G(\lambda + \xi)$ as a power series
in $\xi$:
$$
G(\lambda + \xi) \ = \ G(\lambda) \ + \ {dG \over d\lambda} \xi \ + \ 
{1 \over 2} {d^2G \over d\lambda^2} \xi^2 \ + .... \ + \ 
{1 \over n!} {d^nG \over d\lambda^n}\xi^n  \ + ....
\eqno (22)
$$
At this point it is convenient to
specialize the discussion to the case in which $G(\lambda)$
is a Gaussian with width $\sigma$, namely
$$
G(\lambda) \ = \ \left ( {1 \over \sqrt \pi \sigma} \right ) ^{1/2}
\ \exp [- {(\lambda - \lambda_0)^2
\over 2 \sigma^2}] \ = \ \SH_0(\lambda, \sigma) \ .
\eqno (23)
$$
In this case the derivatives in Eq. (22) are of a simple form:
$$
{d^nG \over d\lambda^n} \ = \ \left ( {1 \over \sqrt \pi \sigma} \right )
^{1/2} \sigma^{-n} \ \exp
[-{(\lambda - \lambda_0)^2 \over 2 \sigma^2}] \ R_n({\lambda - \lambda_0
\over \sigma}) \ \ ,
\eqno (24)
$$
where the $R_n$ are polynomials of order $n$ in $(\lambda - \lambda_0)
/ \sigma$.
Using the relations given by Abromowitz \& Stegun (1972), the
$R_n$ may be rewritten in terms of a sum of Hermite polynomials $H_n$
of equal and smaller order.
Invoking the definition (6) of the Hermite functions $\SH_n$,
one obtains at last
$$
G(\lambda + \xi) \ = \ \sum \limits _{n=0} ^\infty {1 \over n!}
\left ( {\xi \over \sigma } \right )^n
\ \sum \limits _{j=0} ^n E_{nj} (2^j j!)^{1/2} \ \SH_j(\lambda, \sigma) \ ,
\eqno (25)
$$
where the first few rows of the matrix $E_{nj}$ are given by
$$
E_{nj} \ = \ \pmatrix{
1 \cr
 &-1/2\cr
-1/2& &1/4\cr
 &3/4& &-1/8\cr
3/4& &-3/4& &1/16\cr
 &-15/8& &5/8& &-1/32\cr} \ \ ,
\eqno (26)
$$
and where, for clarity, zeros are represented by blanks.

Since $j$ never exceeds $n$ in the sums in Eq. (25),
$(2^j j!)^{1/2}/n!$ decreases rapidly with $j$.
Thus, so long as $\xi < \sigma$,
the coefficients multiplying
$\SH_j$ decrease rapidly as well.
Suppose that the total range of $vc/\lambda_0$ that occurs on the
stellar disk is smaller than $2 \sigma$.
Then the factor $G$ in the
integrand of (Eq. 20) is everywhere dominated
by low-order Hermite functions.
In particular, one can expect $h_3$ to be the largest contributor
to the line asymmetry.
Barring a fortuitous cancellation that makes $h_3$ nearly zero,
it should therefore dominate the integral, and
the shape of the line bisector variations
should be essentially logarithmic in the line depth,
as shown in the previous subsection.
This result is no surprise;
high-order Hermite functions correspond in a rough way to
high-frequency Fourier components.
These components are attenuated by convolution with the
intrinsic line profile.
Only when the range of velocities generated by rotation and
pulsation becomes larger than the broadening width characteristic
of the intrinsic line profile can the shape of the bisector
begin to reflect the spatial distribution of those velocities.

In the case of 51 Peg, the central part of the line
$\lambda$6253 Fe I as it would be observed with infinite spectral resolution
is well approximated by a Gaussian
with FWHM = 8.5 km s$^{-1}$,
corresponding to $\sigma = 3.61$ km s$^{-1}$.
The best current estimate for $v \sin i$ for 51 Peg
is 2.35 km s$^{-1}$ (Hatzes et al 1997).
It follows that for surface oscillation amplitudes
$V_p k \ell \leq 1.2$ km s$^{-1}$, the shape
of observed line profile variations should be as in Eq. (15).
The numerical simulations we describe below yield this
result, whereas in those of Hatzes (1996) the bisectors show a more
complicated structure.
The agreement between this analytical
treatment and the numerical one we present below bolsters our
confidence that the latter has been carried out correctly.
As a further check, we have compared a sample of our simulated
line profiles with those computed by Schrijvers (1997)
using a different code, and we find good agreement.

Real stellar line profiles are not Gaussian;
they typically have cores that are well approximated
by Gaussians, but their wings are wider.
The principal effect of this departure from the simple
shape leading to Eq. (15) is to change the line profile's
derivatives appearing in the denominator of Eq. (13),
but only for the line wings.
But the shallow parts of the line wings are seldom used for
line bisector analysis, because these parts of the profile
are sensitive to blends with other spectrum lines.
For the relative line intensities $U$ actually used, simulations show that
the departures of the observed line profiles from Gaussian
shapes are unimportant.

\vskip12pt
\parindent=0pt
{\it B. Numerical Simulations}
\parindent=20pt

The foregoing analysis shows that a representation of the
observed line profiles in terms of Hermite polynomials is likely
to be useful.
Moreover, it gives some idea of the line shape behavior to be expected 
in slowly-rotating stars.
To interpret observed line profiles in a quantitative fashion,
however, we must perform numerical simulations of the shape of
the line profile and its variations during a cycle of pulsation.
Given the variation of the line intensity with wavelength and with
oscillation phase,
one can treat the simulated line profiles as if they were real
data, computing the variation with phase of any desired parameter
of line shape or position (wavelength of the line center-of-gravity,
the span or curvature of the line bisector, the magnitudes of
any of the Hermite coefficients $h_i$, {\it etc.}).
The amplitudes and phases of the oscillating part of any of these
parameters are the fundamental quantities that we use for comparison
with observations.

A significant complication is introduced by our lack of knowledge
of the characteristics of the pulsation modes that may exist on
51 Peg or similar stars.
In particular, we know nothing of the likely mode vertical
velocity amplitude $V_p$,
nor of $\ell$, $m$, or the inclination $i$.
Despite the arguments presented in section II,
it is not obvious even that the pulsating flows must 
have large $k$, and hence appear essentially
horizontal.
For instance, a periodic variation in the correlation between velocity
and intensity in granulation would cause a time-dependent convective
blue shift,
and seemingly vertical oscillating flows
could result. 
Experience with the Sun offers no useful guides regarding pulsations
with such large amplitudes, so we must consider all possibilities
and hope that the observations will help us to choose among them.

For the purposes of this paper, we have computed the line profile variations
for all horizontal eigenfunctions $Y_{\ell}^m$ satisfying 
$1 \leq \ell \leq 10$ and $0 \leq m \leq \ell$,
for inclinations $i = \{15 \degree,\  45\degree, \ 75\degree\}$.
We used two combinations of $k$ and $V_p$;
$k=0$, $V_p=100$ m s$^{-1}$, corresponding to purely vertical flows,
and $k=100$, $V_p=1$ m s$^{-1}$, corresponding to flows that are
predominantly horizontal.
We assume that the response to oscillations is sufficiently linear
that different amplitudes may be simulated simply by scaling
these two small-amplitude cases.
This assumption is not strictly correct, even for the relatively small
oscillating velocities (1 km s$^{-1}$) that we consider below (Schrijvers
{\it et al.} 1997)
Nonlinear effects are small at these amplitudes, however,
and their presence would not alter our conclusions.
Finally, in order to make comparisons with observations of both 51 Peg
and $\tau$ Boo, we used $v \sin i$ values of 2.4 and 15 km s$^{-1}$.
The final grid contained 390 models for each star,
implying a fairly large amount of computation.
A more significant difficulty, however, is that for any given
observable line parameter, one can usually find particular values
of $\ell$, $m$, and $i$ for which the oscillating
component of that parameter nearly vanishes because of cancellation
of the signal over the visible disk of the star.
Other line parameters need not vanish at the same time, and
usually do not.
Thus, the ratio of the oscillating components of two different 
observable line parameters 
(center-of-gravity wavelength
and line bisector curvature, for instance) may
in principle take on any value between zero and infinity,
even though the typical value of the ratio may be well defined.
This is unfortunate, since such ratios are the natural diagnostics
to use when comparing observations of different kinds.
To accommodate this problem, we present the results of our
computations of oscillating amplitude in the form of plots
that show, as a function of $\ell$, the minimum, mean, and
maximum value of the plotted parameter as $m$ and $i$ are
allowed to run over the ranges given above.
For most of our conclusions we shall rely upon the mean values;
we note that these conclusions are not significantly changed
by using median values rather than means.
But it is important to realize that statements about relative
values of observable parameters hold only in a statistical sense,
and that discrepant ratios (either large or small) can easily be found.

The first step in computing emergent line profiles is to evaluate
the line-of-sight component of the sum of the rotational and oscillating
velocities on a suitably fine grid in space and time.
We assumed solid-body rotation ({\it i.e.}, angular velocity independent
of latitude), and oscillation eigenfunctions as in Eq. (2,3,4),
with the spherical harmonic coordinate $z$-axis aligned with the
rotation axis.
We used a spatial grid of 201 x 201 points, with uniform separation
of 0.01 $R_*$.
The number of spatial points lying within the stellar disk was
therefore about 31400.
Figure 3 illustrates line-of-sight velocity contours for a variety
of circumstances.
In the absence of limb darkening, the area between adjacent contours
would be proportional to the Doppler shift density $W(v)$ in Eq. 21.
Note that for small $k$, the oscillatory motions appear largest near
the center of the stellar disk, while for large $k$ they are largest
near the limbs.
One effect of this tendency is that, for large $k$ (horizontal flows), 
the oscillating
velocities are dominated by small regions near the East and West limbs,
even for fairly large $\ell$.
In contrast, small-$k$ (vertical) flows have substantial amplitude over
a larger area near disk center.
For this reason, the cancellation that prevents detection of high-$\ell$
pulsations is less effective for horizontal flows than for vertical ones.

To evaluate the emergent line shape, each spatial point on the stellar
disk is assumed to contribute an intrinsic line profile whose strength depends
upon $\mu$ (the cosine of the angle between the local normal to
the stellar surface and the line of sight),
and whose center wavelength depends upon the local line-of-sight component
of velocity.
For the computations reported here, we 
took the line strength to be proportional to $(1+\mu)/2$,
assumed the intrinsic line shape
to be independent of position on the disk (a fairly good approximation
for most lines)
and finally assumed that it could be represented by 
the sum of two Gaussians with widths differing
by a factor of 2.5 and central depths differing by a factor of 10.
The narrower of these Gaussians had FWHM = 8.5 km s$^{-1}$
($\sigma = 3.61$ km s$^{-1}$).
These profiles are a fairly close approximation to those used
in the simulations by GH
(see Figure 4).
We also tried profiles consisting of the narrow Gaussian alone.
Using these profiles with truncated wings made 
no significant difference in our
results, because
the affected parts of the line profiles
lie at fractional line intensities
that are too shallow to enter into the bisector analysis.
For these reasons we think that the 2-Gaussian computations
give an adequate representation of the oscillating parts of the line
profiles, and minor changes in the choice of line profiles
(or in their shape variation across the stellar disk)
would not alter our conclusions in any important way.

We computed line profiles on a wavelength grid with a spacing of 0.1
AFOE pixels, corresponding to approximately 0.003 \AA,
or about 200 m s$^{-1}$.
To insure adequate sampling of the oscillation, 
we computed line profiles at 20 phase
points equally spaced throughout the cycle.
Figure 5 illustrates some samples of such line profile time series.
A large oscillating amplitude (maximum surface velocities of
4 km s$^{-1}$) has been
assumed in this Figure, in order to produce easily visible distortions
of the profiles for $v \sin i = 15$ km s$^{-1}$.
But even for this large amplitude, the profile changes when $v \sin i =$2.4
km s$^{-1}$ are subtle,
consisting almost entirely of modest variations in line width, depth,
and displacement.

The observed line shapes differ from those just described because
they suffer additional smoothing by an instrumental point spread function.
In order to allow comparisons between our AFOE data and those
described by GH, we computed three related sets
of line profiles: (1) unsmoothed, {\it i.e.} exactly the
results just described; (2) smoothed with a Gaussian with
FWHM = .047 \AA, corresponding to spectral resolution $R = 10^5$;
(3) smoothed with the observed PSF of the AFOE, as determined from
a Th-Ar spectrum taken during October 1995, giving $R \simeq 50000$.
Figure 6 shows the bottom-most line profiles from Fig. 5
for the 3 different amounts
of instrumental smoothing and for $v \sin i =$ 2.4
and 15 km s$^{-1}$.
The spectral resolution makes a noticeable
difference for small $v \sin i$, but is less important
for higher rotation speeds.

Having computed a time series of line profiles that have
been distorted by rotation and pulsations and smoothed by
convolution with an instrumental line profile,
we next estimated the various line shape parameters
for the profiles obtained at each time step.
To estimate line bisectors, we used linear interpolation
between computed wavelength points to estimate the
wavelength separation between points at specified relative
intensities in the profile.
In this way, we computed bisector positions at 1\% 
intervals for relative line intensities between 5\% and 98\%.
We found that linear interpolation led to smoother bisectors
than did higher-order schemes,
but even so, the bisectors displayed small-scale oscillations
that added unnecessary noise.
To filter out this noise, we fit each bisector with a 7th-order
polynomial in the relative line intensity, and used this polynomial
as a representation of the bisector.
To derive bisector span and curvature values, we evaluated the
bisector displacement $B$ from line profiles smoothed to $R=100000$,
at relative line
intensities
of $d_1 = 25$\%, $d_2 =$48\%, and $d_3 =$87\%.

To estimate the parameters in the Hermite function expansion,
we used a Levenberg-Marquardt iterative algorithm
(Press {\it et al.} 1987) to perform a least-squares fit of
the function described in Eq. (9) to the
$R=50000$ computed profiles.
Experience suggested that Hermite terms higher than $h_5$ were
likely to be unobservably small in real data, so we truncated
the expansion at that point.
Also, the computed profiles were always normalized so that their
continuum level was unity.
We arbitrarily set the flux ratio between line center and the continuum
at 0.7 for the case of $V_p = 0$ and instrumental resolution
$R = \infty$.
The parameters being fit then consisted of $D_j$, $\lambda_{cj}$, $\sigma$,
and $h_3$, $h_4$, $h_5$ from Eq. (9).
Combined with the bisector measures $S_b$ and $C_b$, these
parameters provided a variety of distinct (though not all
independent) measures of the line shape.
Notice that the Hermite coefficient $h_4$ measures a symmetric
distortion of the line profile that is distinct from the Gaussian
width $\sigma$;
this piece of information is missing from the bisector analysis,
which responds only to antisymmetric distortions of the profile.
Figure 7 shows time series of all 8 of these parameters for
pulsational motions that are essentially horizontal:
$v \sin i = 2.4$ km s$^{-1}$, $\ell = m = 4$, $V_p = 1$ m s$^{-1}$,
$k = 100$, and $i = 75 \degree$.
Also shown on the figure are sinusoidal fits to the variations,
assuming a period equal to the oscillation period.
The variations observed are all of this character,
with at most a small admixture of harmonics.
It is therefore reasonable to describe the computed variations
in terms of the amplitude and phase of this principal sinusoidal
component.
In what follows we shall attend primarily to the amplitudes
(whose values are displayed in Fig. 7),
paying little attention to the phases.

We note in passing that many pulsating stars show line profile variations
that are distinctly non-sinusoidal, with the second and higher harmonics
of the principal period having significant amplitude ({\it e.g.,} 
Schrijvers {\it et al.} 1997).
This may occur if $v \sin i$ is greater than the intrinsic line width
$\sigma$, and if the horizontal gradients of 
line-of-sight pulsational velocity on the stellar
surface are comparable to those 
arising from rotation.
Even if the pulsation velocities are purely vertical,
their horizontal gradients scale with $\ell$.
Thus, for modes with $\ell$ of only a few on stars with sufficient $v \sin i$,
one may observe quite nonlinear behavior even when $V_p$ is
much less than the surface sound speed.
These considerations are irrelevant for 51 Peg, because of its small
$v \sin i$, but they start to become significant for $\tau$ Boo.

At this point it is useful to recall the results of the previous
section and ask how the bisector variations computed for the
case shown in Fig. 8 agree with expectations from the analytical treatment.
Since $V_p + v \sin i$ is slightly smaller than 
$\sigma$, one expects the bisector variation
to be described by a constant displacement plus a logarithmic
function of relative line depth.
A function of this form is shown in Figure 8 for the oscillation
phase $\phi=0$.
The agreement with expectation is good,
even though the assumptions leading to Eq. (15) are only
marginally satisfied.
Also, the ratio of bisector span to curvature amplitude 
and of bisector span to $h_3$
are in good agreement with those estimated in Eqs. 17-18, as they
must be if the relative line intensity
variation of the bisector position is essentially
logarithmic.
This agreement verifies the close connection between $h_3$ and
variations of bisector shape, at least in slowly rotating stars.
However, it immediately raises a conflict with the models described
by GH.
Those models were constructed to match 51 Peg's observed radial velocity
amplitude of 56 m s$^{-1}$;
the corresponding amplitudes computed for bisector curvatures 
lay between 25 and 56 m s$^{-1}$, depending upon the choice of $k$
and $i$.
Thus, the models by GH yielded bisector curvature amplitudes that
were similar to the radial velocity amplitudes, with the ratio of amplitudes
between 0.45 and 1.0, depending on the specific parameter choices.
This is not so for the mode parameters illustrated in Fig. 7,
where the bisector curvature amplitude is only 0.09 as large as the
radial velocity amplitude.
If one considers a range of inclinations and $m$ values for $\ell = 4$,
it develops that the typical ratio of bisector curvature to velocity
is slightly smaller yet -- about 0.07.
This behavior persists at all $\ell$ values.
Because of our models' small bisector curvatures, it is not possible for
them to reproduce the observations reported by GH,
except perhaps in a few special cases.
These cases occur when the geometry happens to cause a nearly exact
cancellation of the pulsating velocity signal, but not of the bisector
curvature.
A major part of the disagreement between our results and those
by GH derives from this lack of accord between the fundamental features
of our models, and from the consequent ambiguity about how different
observables should be related to one another.
We feel obliged to point out this disagreement;
the remainder of this discussion will, however, proceed on the assumption
that our simulations are correct.

Figures 9 to 12 display the results for our complete set of line profile
simulations.
Each plot shows the amplitude at the pulsation frequency for each of
several observable quantities, plotted against $\ell$.
The 4 plots show the possible combinations of small and large $k$
($k=0$, $V_p=100$, and $k=100$, $V_p=1$) with small and large
$v \sin i$ ($v \sin i = 2.5$ and $v \sin i = 15$ km s$^{-1}$).
For each $\ell$ we display the minimum, maximum, and mean values of
each plotted observable, with the statistics taken over the
$\ell +1$ possible $m$ values (only one sign of $m$ being necessary
for the purpose of computing amplitudes), and 3 values
of the inclination $i$.
Unless otherwise noted, the following discussion will focus on the
behavior in typical cases, {\it i.e.,} those described by
the ``mean'' curves.

Several general features are apparent in Figs. 9-12.
The magnitude of observed line profile changes tends to decline
at high $\ell$, because of increasing amounts of cancellation
across the stellar disk.
This tendency is much more pronounced for vertical than for horizontal
velocities,
both because the horizontal velocities scale with $\ell$ ($V_p$ being
held constant), and because the horizontal velocity signal is dominated
by contributions from small and foreshortened regions near the stellar
limbs.
The decrease in amplitude with $\ell$ is also more prominent for small
than for large $v \sin i$, because in the latter case cancellation
occurs only over small sub-areas of the star, with points within each such area
having similar projected rotational velocity.
Observables that depend upon distortions of the shape of the line
tend to become small also for $\ell$ very small.
This occurs because, at low $\ell$, the dominant effect of surface velocities
is to change the radial velocity averaged over the visible hemisphere,
with the distortions due to velocity dispersion being less important.
Thus, line distortion parameters such as the bisector curvature and $h_3$
have largest amplitude (both in absolute terms and relative to
the radial velocity amplitude) for $\ell$ between 3 and 6,
provided that $v \sin i$ is small.
For $v \sin i = 15$ km s$^{-1}$ (and particularly for horizontal velocities),
the largest amplitudes may be found at larger $\ell$.
Finally, for slow rotation, the shape of the $\ell$-dependent variation
is similar for the bisector span, bisector curvature, and $h_3$.
This reflects the large smoothing effect of the intrinsic line profile.
In these cases the distortion of the line bisector is always
nearly logarithmic in shape;
all that can vary with $\ell$ is the size of the distortion, 
and (somewhat independently) the size of the associated line displacement.

As a rough indication of the magnitudes involved, we assume
that $\ell = 5$ (which gives, on average, the largest values of $h_i$
for a given radial velocity), and
take $v \sin i$ and velocity amplitudes as observed for
51 Peg and to $\tau$ Boo.
The corresponding amplitudes of the other observables
(again assuming the ``mean'' curves) are then given in Table 1.
Note that for 51 Peg, the variations of both the bisectors and
the Hermite coefficients are very small.
In $\tau$ Boo, by contrast, both sorts of measurements should
have easily observable amplitudes.
We shall return to these issues after a discussion of our observations.

\vskip10pt
\centerline {\bf Table 1}
\centerline{Expected Amplitudes of Pulsating Quantities}
$$\vbox
{\halign{#\hfil \qquad & \hfil#\hfil \qquad & \hfil#\hfil \qquad &
\hfil#\hfil \cr
\multispan3\hrulefill \cr
$l=5$, $k=100$& & \cr
\multispan3\hrulefill \cr
 &51 Peg & $\tau$ Boo \cr
\multispan3\hrulefill \cr
$V_{Dop}$ (m s$^{-1}$)& 56.& 468.\cr
Bisector Span (m s$^{-1}$)& 58.9& 2821.\cr
Bisector Curv. (m s$^{-1}$)& 7.5& 978.\cr
$h_3$ (\%)& 0.508& 14.8\cr
$h_4$ (\%)& 0.103& 10.1\cr
$h_5$ (\%)& 0.095& 8.3\cr
\multispan3\hrulefill \cr
}}$$

\vskip20pt
\parindent=0pt
{\bf IV. Observations}
\vskip8pt
{\it A. Procedures}
\parindent=20pt

We obtained spectroscopic observations of 51 Peg and $\tau$ Boo
using the Advanced Fiber Optic Echelle (AFOE) spectrograph,
located at the 1.5 m Tillinghast telescope at the Smithsonian
Institution's Whipple Observatory on Mt. Hopkins, AZ.
The AFOE (Brown {\it et al.} 1994) is a cross-dispersed and fiber-fed
echelle spectrograph designed specifically for the precise
measurement of stellar Doppler shifts.
For the observations described here, we operated the AFOE at a
spectral resolution of $R \simeq 50000$.
To provide a precise wavelength reference for radial velocity
measurements, almost all of the observations reported here were
obtained using an I$_2$ absorption cell,
which imposes a dense spectrum of narrow molecular absorption lines
on the stellar spectrum, mostly within the wavelength range between
500 and 610 nm.

The spectra used in the current study were taken as part of our program
to search for low-mass companions to Sun-like stars.
The observing and data reduction techniques have been described in
an earlier paper (Brown {\it et al.} 1998);
here we shall repeat only their essential features.

The AFOE spectra cover about 55\% of the wavelength range between
395 and 675 nm in 24 diffraction orders.
The band around 625 nm containing the lines analyzed by G97
and by GH falls near the center of order 36, so good comparisons
with their results for these lines are possible (within the limits
imposed by our poorer spectral resolution).
For both stars, our observations 
usually consisted of a sequence of 3 consecutive 10-minute integrations.
Most often we took one such set per night, but on some occasions,
to obtain better sampling of the short radial velocity periods of these
stars, we took 2 sequences, one near the beginning and one near the end
of the night.
The signal-to-noise ratio for single integrations was typically about 150:1
in the continuum near 600 nm.
To observe line profile perturbations with amplitudes of a fraction of
a percent, it was therefore necessary to average together the results
from a few tens of spectrum lines.
Figure 13 shows typical spectra of the region near 625 nm,
for both 51 Peg and $\tau$ Boo.

The data set for 51 Peg consisted of 59 spectra taken on 18 different
nights between 2 Nov 1995 and 26 July 1996.
That for $\tau$ Boo consisted of 90 spectra taken on 23 nights between
25 June 1996 and 26 March 1997.
In both cases our observations tended to cluster around the times
of full moon (the AFOE is used for bright-star spectroscopy, hence
is allocated nights mostly during lunar bright time).
The effects of this periodicity in our observing window function
are a prominent feature in the periodograms of our data.

We used distinct analysis methods to measure Doppler shifts and
line profile distortions.
We estimated Doppler shifts from the relative displacement between
the stellar spectrum and the I$_2$ spectrum,
measured in the diffraction orders (38 to 44) where the I$_2$ lines
are strong.
This process involves generating a model spectrum of the form
$$
I_{obs}(\lambda ) \ = \ [ S(\lambda + d\lambda_s) T(\lambda + d\lambda_i)]
\star P \ \ ,
\eqno (27)
$$
where $S(\lambda)$ is a standard spectrum of the target star obtained
without the I$_2$ cell, $T(\lambda)$ is the transmission of the I$_2$
cell, measured with spectral resolution of $5 \times 10^5$,
$d\lambda_s$ and $d\lambda_i$ are displacements of the star and
I$_2$ spectra relative to their standard positions on the detector,
$P$ is the spectrograph point spread function, and $\star$ denotes
convolution.
For more details of the Doppler analysis, see Noyes {\it et al.} (1997)
and Korzennik {\it et al.} (1997).

To measure line profile shapes, we used an iterative $\chi^2$-minimization
method to fit parameters in 
Eq. (9).
As a first step, we estimated the shift of the I$_2$ spectrum
relative to its standard position,
and formed an estimate of the I$_2$ cell transmission from this
displacement combined with a low-noise spectrum of the flat-field
source shining through the I$_2$ cell.
Dividing the observed spectrum by this estimated I$_2$ transmission
then gave a stellar spectrum from which the I$_2$ spectrum had been
substantially removed.
This compensation process was not perfect, however,
so to avoid misleading results caused by residual I$_2$ features
moving across stellar lines, we excluded from consideration the
7 orders in which the raw I$_2$ spectrum shows line depths greater than
3\%.
Fitting the spectra then proceeded in two stages.
First we fit a model with all parameters free
(including line center wavelengths),
using as input an average of 3 stellar spectra obtained without the
I$_2$ cell.
Then, using this model as a starting guess,
we fit each individual spectrum holding the relative line positions
constant, but allowing displacements of the entire ensemble of lines.
Though including this ensemble displacement
was essential for obtaining an adequate fit to the observations,
it was not useful for measuring radial
velocities because of the absence of a suitable reference spectrum
to monitor instrumental drifts.
We performed independent fits for spectrum segments of roughly 1.5 nm
width (about one-quarter of the wavelength range in a single order),
for each of several orders.
The segments analyzed were chosen to avoid the edges of the orders
(where flat-fielding is problematic),
to contain several separated medium-strength absorption lines,
and to have at least a short stretch of apparently clean continuum
at each end.
We did not include very strong lines in any of these segments.
To combine the results for line profile shape parameters in different
orders, we simply formed their unweighted averages over order.
Figure 14 shows one example of the 51 Peg spectrum segment near 625 nm,
with the fitted model spectrum overlaid.

\vskip 12pt
\parindent=0pt
{\it B. 51 Pegasi}
\parindent=20pt

The results of the above analyses were time series of the stellar
radial velocities for both stars,
and similar time series of the various line depth and shape parameters
from Eq. (9).
Figure 15 illustrates the radial velocity data for 51 Peg plotted against
phase, assuming the RV period to be 4.2312 days
(Marcy {\it et al.} 1997),
and taking the phase zero to be the published time of maximum
radial velocity (JD = 2450203.947).
The agreement between the AFOE velocity period and amplitude
and those of Mayor \& Queloz (1995) and of Marcy {\it et al.}
is evidently very good.
The AFOE data provide an independent second confirmation of the presence
of the 4.23-day RV variation in 51 Peg.

Figure 16 shows the corresponding time series for 
the Hermite coefficients
$h_3$, $h_4$, and $h_5$,
and for the line depth ratio $\lambda$6252 V I to $\lambda$6253 Fe I. 
The data used for the average values plotted here came from orders
35, 36, and 47, with 65 distinct line profiles being fitted.
We chose these particular orders to minimize the time series rms
of the $h_3$ coefficient, when combining the orders in an unweighted average.
The figure also shows the best-fit 4.23-day sinusoid for each observable,
and the corresponding amplitudes.
The zero points of phase for these plots are once again taken to be
the time of maximum radial velocity.
To assess the significance of these signals, one may examine
the periodograms of the time series.
Portions of these are shown in Figure 17, with the RV
frequency and the possible second periodicity (2.575 days) suggested
by GH indicated by vertical lines.
The most important conclusion from the periodograms is that none
of the peaks observed are significant;
the only prominent features are the roughly periodic
sets of peaks separated by 0.03 cycle/day,
resulting from the monthly periodicity of our observing window.
We estimate the 1-standard deviation
uncertainty in the fitted sinusoid amplitudes
as the amplitude corresponding to the average power in a band
of width $\pm$1 cycle/month centered on the nominal frequency.
The measured amplitudes and their uncertainties are tabulated in
the first two rows of Table 2.
The errors exceed the fitted amplitudes in all cases,
so that in all cases we have measured only upper limits for the
associated pulsation amplitudes.

\vskip 12pt
\parindent=0pt
{\it C. $\tau$ Bo\"otis}
\parindent=20pt

Figure 18 shows the time series of Hermite coefficients for 
$\tau$ Boo, and Figure 19 shows the corresponding periodograms.
We averaged the Hermite coefficient time series
over echelle orders 35, 40, 46, and 49, modeling 57 lines.
For this star we single out only one frequency, namely that of
the RV variation.
We do not compute the line-depth ratio in this case,
since in this relatively hot star the $\lambda$6253 V I line is too weak to 
be used in a meaningful ratio.
Estimated amplitudes and uncertainties once again appear in Table 2.
The derived amplitudes are once again not significant,
but it is worth noting that this star does show evidence for weak,
broad-band line profile variations, particularly in $h_3$ at the
radial velocity frequency $\pm$ 0.05 cycles/day.
This variability may be associated with evolving magnetic
activity combined with rotation,
since the rotation period of $\tau$ Boo is thought to be identical to
its radial velocity period.
Such line profile variations may also explain the excess noise
reported in the radial velocity time series of $\tau$ Boo by
Butler {\it et al.} (1997)

\vskip10pt
\centerline{\bf Table 2}
\centerline{Observed Amplitudes and Errors for 51 Peg \& $\tau$ Boo}
$$\vbox
{\halign{\hfil#\hfil \qquad & #\hfil \qquad & #\hfil \qquad &
#\hfil \cr
\multispan4\hrulefill \cr
  &51 Peg & 51 Peg & $\tau$ Boo \cr
  & 4.231 d & 2.575 d & 3.313 d \cr
\multispan4\hrulefill \cr
$h_3$ (\%) & 0.147 $\pm$ 0.26 & 0.344 $\pm$ 0.38 & 0.270 $\pm$ 0.35 \cr
$h_4$ (\%) & 0.272 $\pm$ 0.42 & 0.238 $\pm$ 0.45 & 0.148 $\pm$ 0.30 \cr
$h_5$ (\%) & 0.219 $\pm$ 0.19 & 0.323 $\pm$ 0.19 & 0.190 $\pm$ 0.17 \cr
Line Ratio & 0.277 $\pm$ 0.30 & 0.416 $\pm$ 0.28 &\quad  -- \cr
\multispan4\hrulefill \cr
}}$$

\vskip 20pt
\parindent=0pt
{\bf V. Discussion}
\parindent=20pt

To make sense of the observed limits on line profile variations
contained in Table 2
we must compare them with the variations that would be expected
from NRPs,
choosing the pulsation parameters to be consistent (insofar as
possible) with observations of radial velocity and
of line bisector fluctuations.
Because of the aforementioned disagreement between our models
and those by GH, there are several ways to do this
comparison.
For 51 Peg we adopt the  approach of normalizing the ``mean''
plots in Fig. 9-12 to give the observed amplitude for each one of the
observables for which variation has been reported: radial velocity,
bisector span, and bisector curvature.
We may also choose a normalization to match the pulsation amplitude
$V_p$ in one or more of the models described by GH as fitting
their observations.
If their choices for $m$ and $i$
give typical ratios of velocity to line shape parameters,
then the normalization to radial velocity
will give correct amplitudes for $h_3$, $h_4$, and $h_5$.
If those values of $m$ and $i$ result in atypical ratios
between the observable quantities, or if our computed relationship
between radial velocity and the $h_j$ is in error for some other reason,
then normalizing to the bisector span or curvature
should at least give a result that is most nearly comparable to
the corresponding observations by G97 or by GH.
No such normalizations are necessary for interpretation of
the line depth ratios;
in that case G97 and the AFOE spectra are measuring the same thing,
albeit with different spectral resolution.
We compare expected and observed Hermite coefficient amplitudes
only for $\ell=5$, since that value of $\ell$ gives (on average)
the largest variation in Hermite coefficients,
bisector span, and bisector curvature for a given radial velocity
amplitude.

We take the reported amplitudes of oscillation of the various observables
to be as follows:
radial velocity, 56 m s$^{-1}$; bisector span (measured from Fig. 2
of G97), 38 m s$^{-1}$; bisector curvature (from GH), 45 m s$^{-1}$;
line depth ratio 6252 V I to 6253 Fe I (also measured from Fig. 2 of
G97), 0.6\%.
As an explicit illustration of the process we followed, consider 
first normalizing to radial velocity, assuming that $v \sin i = 2.4$
km s$^{-1}$ and $k = 100$.
Figure 9 shows that for $V_p = 1$ m s$^{-1}$, the mean detected
radial velocity is 3.42 m s$^{-1}$, so to produce an observed radial
velocity of 56 m s$^{-1}$ we must multiply $V_p$ by 16.4.
Scaling the Hermite coefficient amplitudes by the same factor,
we obtain expected amplitudes $h_3 = 0.508$\%, $h_4 = 0.103$\%,
and $h_5 = 0.095$\%.
Finally, we express these amplitudes in terms of the 1-$\sigma$ noise
limits derived from the 51 Peg observations, listed as the uncertainties
in Table 2.
The resulting ratios, listed in the first row of Table 3, are a measure
of the significance of our non-detection of a pulsation in the
Hermite coefficients.
For this normalization, $h_3$ should have been seen to oscillate
with an amplitude
roughly twice its nominal detection limit;
the amplitudes of $h_4$ and $h_5$ would be factors of about 4
and 2 below their respective detectability thresholds.
The next three rows of Table 3 show 
the results of identical calculations, but normalized to the reported
amplitudes of the bisector curvature, bisector span,
and pulsation amplitude $V_p$.
The latter refers to the GH model with $\ell=4$, $m=4$, $k=100$, $i=40\degree$,
and $V_p=8.8$ m s$^{-1}$, which they calculate to yield a 
radial velocity amplitude of 56 m s$^{-1}$ and a
bisector curvature amplitude of 32 m s$^{-1}$.
The table, however, refers to the results of our own modeling, which
for the same oscillation parameters yields amplitudes for the
radial velocity
of 37 m s$^{-1}$ and 
for the bisector curvature of 3.8 m s$^{-1}$.
Last, the table shows the model-independent ratio between
G97's reported line depth ratio and 
our detection limit for the same quantity.

\vskip10pt
\centerline{\bf Table 3}
\centerline{Significance of Non-Detection of Pulsation (Units of $\sigma$)}
$$\vbox
{\halign{#\hfil \qquad & #\hfil \qquad & #\hfil \qquad &
#\hfil \qquad& #\hfil \qquad \cr
\multispan5\hrulefill \cr
Normalization & $h_3$ & $h_4$ & $h_5$ & VI/FeI \cr
\multispan5\hrulefill \cr
  &{\bf 51 Peg} &  &   & \cr
\multispan5\hrulefill \cr
$V_{Dop}=56$ m s$^{-1}$ & 1.95 & 0.24 & 0.50 & \cr
Curv=45 m s$^{-1}$ & 9.78 & 1.47 & 2.99 & \cr
Span=38 m s$^{-1}$ & 1.06 & 0.16 & 0.32 & \cr
$V_p=9$ m s$^{-1}$ & 1.94 & 0.28 & 0.48 & \cr
Line Ratio =0.6\% &  &  &  & 2.14 \cr
\multispan5\hrulefill \cr
 &{\bf $\tau$ Boo} & & & \cr
\multispan5\hrulefill \cr
$V_{Dop}=468$ m s$^{-1}$ & 42.3 & 33.7 & 48.8 & \cr
\multispan5\hrulefill \cr
}}$$

Table 3 leads us to two conclusions.
First, with high confidence
we may reject the presence in our data of
line profile variations of the size and form reported by G97 and by GH.
This follows from the close connection between the line bisector span and
curvature indices and our $h_3$ line shape parameter.
This connection applies
so long as the range of radial velocities seen on the disk of the
star is smaller than the intrinsic width of the stellar absorption
lines, which is clearly the case for 51 Peg.
Hence, if there exists a bisector curvature variation 
as reported by GH, then we should see
a corresponding $h_3$ variation, independent of the mechanism responsible.
We do not.
Moreover, we get our most significant non-detection (9.8$\sigma$) when
we normalize to the reported amplitude of the
bisector curvature signal, which in turn is GH's best evidence
for a dynamic atmospheric phenomenon.
Second, we conclude that
51 Peg's time varying velocity signal could nevertheless
result from NRPs;
the changes in line shape corresponding to the observed radial
velocities are probably not 
detectable, either by our techniques or in the existing line
bisector observations.
This conclusion follows from the relatively small line shape
changes that we compute for given velocity amplitudes,
more or less independent of the choice of mode $\ell$ of $m$,
or of the viewing geometry.
The difference between this conclusion and that drawn
by GH arises entirely from the difference between our respective modeling
of the line profile variations resulting from given pulsational motions.

In summary, we find no verifiable evidence supporting nonradial
pulsations as the cause of the radial velocity variations in 51 Peg.
There is certainly no such evidence in our data, and if the
line profile variations reported by G97 and by GH were
in fact present, then we should have detected them.
Our simulations show that pulsations 
nevertheless cannot be excluded as the cause of 51 Peg's radial
velocity variation, since
the line profile changes corresponding to a radial velocity signal
of 56 m s$^{-1}$ may be too small to detect with existing observations.
This ambiguous situation can only be truly resolved by more and
better observations of the line shapes in 51 Peg's spectrum.

To address the larger question of the existence of jovian-mass
objects in small orbits, we may however
turn to $\tau$ Boo,
where the presence
of nonradial pulsations may be investigated more decisively.
Again, we find no evidence for pulsations at the radial velocity
period, though for this star the typical confidence level for non-detection
is quite high: 7$\sigma$ to 35$\sigma$, depending on which
Hermite coefficient is considered.
For $\tau$ Boo, an orbiting companion is a tenable explanation
for the radial velocity variation, while nonradial pulsation is not.

Other explanations of $\tau$ Boo's radial velocity signal seem
to be ruled out, mostly because of the large amplitude of the velocity signal.
In particular, stellar activity combined with rotation would
imply photometric variations of several percent, two orders of
magnitude larger than the photometric stability measured by Baliunas
{\it et al.} (1997).
Moreover, such a mechanism is probably inconsistent with
the RV phase stability that has already been measured in this short-period
system.
Modulation of the line bisector position by some (perhaps magnetic)
mechanism acting on the stellar granulation is implausible,
since the peak-to-peak RV amplitude is similar in magnitude to the
entire convective blue shift caused by the granulation
({e.g.}, Dravins, Lindgren \& Nordlund 1981).

Finally, the absence of line profile variations argues against
the idea that stellar pulsations might be excited by the tidal
interaction with a massive planet.
As GH noted, several elementary considerations make this suggestion 
unpersuasive.
But we are now able to make a more direct observational statement:
because of its small orbital radius and large mass, the inferred
companion to $\tau$ Boo should be the most effective of any of
the claimed extra-solar planets at inducing tidal effects.
Such are not apparent in $\tau$ Boo, so they are almost
certainly negligible in 51 Peg.
The only caveat to this argument is that $\tau$ Boo's rotation is thought
to be tidally synchronized, whereas 51 Peg's is not.

As we mentioned in the Introduction, the hypothesis that Jupiter-mass
objects reside in very small orbits around Sun-like stars
raises problems, notably how such large objects could form
in or migrate into their current positions.
On the other hand, the absence of detectable line-profile shape variations
in $\tau$ Boo, combined with its observed photometric stability
seems to leave no alternative -- some process leading
to hot Jupiters does indeed operate.
This being so, from the perspective of extra-solar planet studies,
the possibility of nonradial pulsations in 51 Peg becomes both
less likely and less interesting.
It is less likely because the likelihood that
a planet causes the velocity signal is greater.
It is less interesting because the existence of such extra-solar planets
is not at issue.
Pulsations are still interesting from the perspective of asteroseismology,
for which large-amplitude pulsations in Sun-like stars would be
a discovery of the greatest importance.
But the pulsation hypothesis has formidable theoretical difficulties
(how to drive modes in Sun-like stars to the implied large amplitudes,
how to select just one high-order overtone for excitation,
and why only a small fraction of Sun-like stars should show such
behavior).
Before accepting that such pulsations exist, unambiguous evidence
will therefore be necessary.
We do not think that the existing evidence for pulsations meets this
standard, especially in view of our inability to find any confirmation
for the line shape or strength signals reported by G97 and GH.
Because of inadequate sensitivity,
the observations of 51 Peg to date are not conclusive, and hence
a continuing investigation of this star with high resolution,
high S/N spectroscopy is warranted.
But until such an investigation is complete, the presence of nonradial
pulsations in this star will remain an unsettled question,
and assertions that a planet does not exist around this star
will remain premature.

We are grateful to the rest of the AFOE team 
(Martin Krockenberger and Adam Contos) and to
the staff at SAO's Whipple Observatory (Bastian van't Saant,
Ted Groner, Perry Berlind, Jim Peters, and Wayne Peters)
for their assistance in obtaining and reducing
the observations described here.
We thank Coen Schrijvers for providing test cases against which to
compare our line profile simulation code,
and Artie Hatzes for many useful discussions.

\vfill\eject
{\bf References}
\def\ref{\leftskip20pt \parindent-20pt \parskip4pt}

\ref
Abramowitz, M. \& Stegun, I. 1972, {\it Handbook of Mathematical Functions},
Dover, New York, p. 775

\ref
Baliunas, S.L. , Henry, G.W., Donahue, R.A., Fekel, F.C., \& Soon, W.H. 1997,
ApJ Lett. 474, L119

\ref
Balona, L.A. 1986, MNRAS 219, 111

\ref
Boss, A.P. 1995, Science, 267, 360

\ref
Brown, T.M. \& Gilliland, R.L. 1994, Ann. Rev. Astron. Astrophys. 32, 37

\ref
Brown, T.M., Noyes, R.W., Nisenson, P., Korzennik, S., \& Horner, S. 1994, 
PASP 106, 1285

\ref
Brown, T.M., Kotak, R., Horner, S.D., Kennelly, E.J., Korzennik, S.,
Nisenson, P., \& Noyes, R.W. 1998,
ApJ Lett (submitted)

\ref
Butler, R.P. \& Marcy, G.W. 1996,
ApJ Lett. 464, L153

\ref
Butler, R.P., Marcy, G.W., Williams, E., Hauser, H., \& Shirts, P. 1997,
ApJ Lett. 474, L115

\ref
Cochran, W.D., Hatzes, A.P., Butler, R.P., \& Marcy, G.W. 1997,
ApJ 483, 457

\ref
Dravins, D., Lindgren, L., \& Nordlund, \AA 1981,
A\&A 96, 345

\ref
Fr\"olich, C. \& Andersen, B.N. 1995, {\it Fourth SOHO Workshop:
Helioseismology}, J.T. Hoeksema, V. Domingo, B. Fleck, \& B. Battrick,
eds, ESA SP-376, v. 1, p. 137

\ref
Gautschy, A. \& Saio, H. 1995, Ann Rev. Astron. Astrophys. 33, 75

\ref
Gray, D.F., Baliunas, S.L., Lockwood, G.W., \& Skiff, B.A. 1992, ApJ 400, 681

\ref
Gray, D.F. 1997, Nature, 385, 795

\ref
Gray, D.F. \& Hatzes, A.P. 1997,
ApJ (in press)

\ref
Hatzes, A.P. 1996, PASP 108, 839

\ref
Hatzes, A.P., Cochran, W.D., \& Johns-Krull, C. 1997, ApJ (submitted)

\ref
Kambe, E. \& Osaki, Y. 1988, PASJ 40, 313

\ref
Korzennik, S., Nisenson, P., Noyes, R.W., Jha, S., Krockenberger, M.,
Brown, T., Kennelly, E. \& Horner, S. 1997, in
{\it Cool Stars 10}, R. Donahue \& J. Bookbinder, eds, ASP Conference
Series (in press)

\ref
Kurtz, D.W. 1990, Ann Rev Astron. Astrophys. 28, 607

\ref
Latham,, D.W., Mazeh, T., Stefanik, R.P., Mayor, M., \& Burki, G. 1989,
Nature, 339, 38

\ref
Marcy, G.W. \& Butler, R.P. 1996,
ApJ Lett. 464, L147

\ref
Marcy, G.W., Butler, R.P., Williams, E., Bildsten, L., Graham, J.R.,
Ghez, A.M., \& Jernigan, J.G. 1997, ApJ 481, 926

\ref 
Mayor, M. \& Queloz, D. 1995, Nature, 378, 355

\ref
Noyes, R.W., Jha, S., Korzennik, S., Krockenberger, M., Nisenson, P.,
Brown, T.M., Kennelly, E.J., \& Horner, S.D. 1997, ApJ Lett. 483, 111

\ref
Press, W.H., Flannery, B.P., Teukolsky, S.A., \& Vetterling, W.T. 1987,
Numerical Recipes: The Art of Scientific Computing,
Cambridge University Press, Cambridge

\ref
Schrijvers, C. 1997, private communication.

\ref
Schrijvers, C., Telting, J.H., Aerts, C., Ruymaekers, E., \& Henrichs, H.F.
1997, A\&A Supp. 121, 343

\ref
Tassoul, M. 1980, ApJ Suppl. 43, 469

\ref 
Toutain, T. \& Fr\"ohlich, C. 1992, A\&A 257, 287

\ref
Vogt, SS. \& Penrod, G.D. 1983, ApJ, 275, 661

\vfill\eject

\hoffset=-0.25in \voffset=-0.5in
\hsize 6.0in
\vsize 8.00in

\input psfig
\psfig{figure=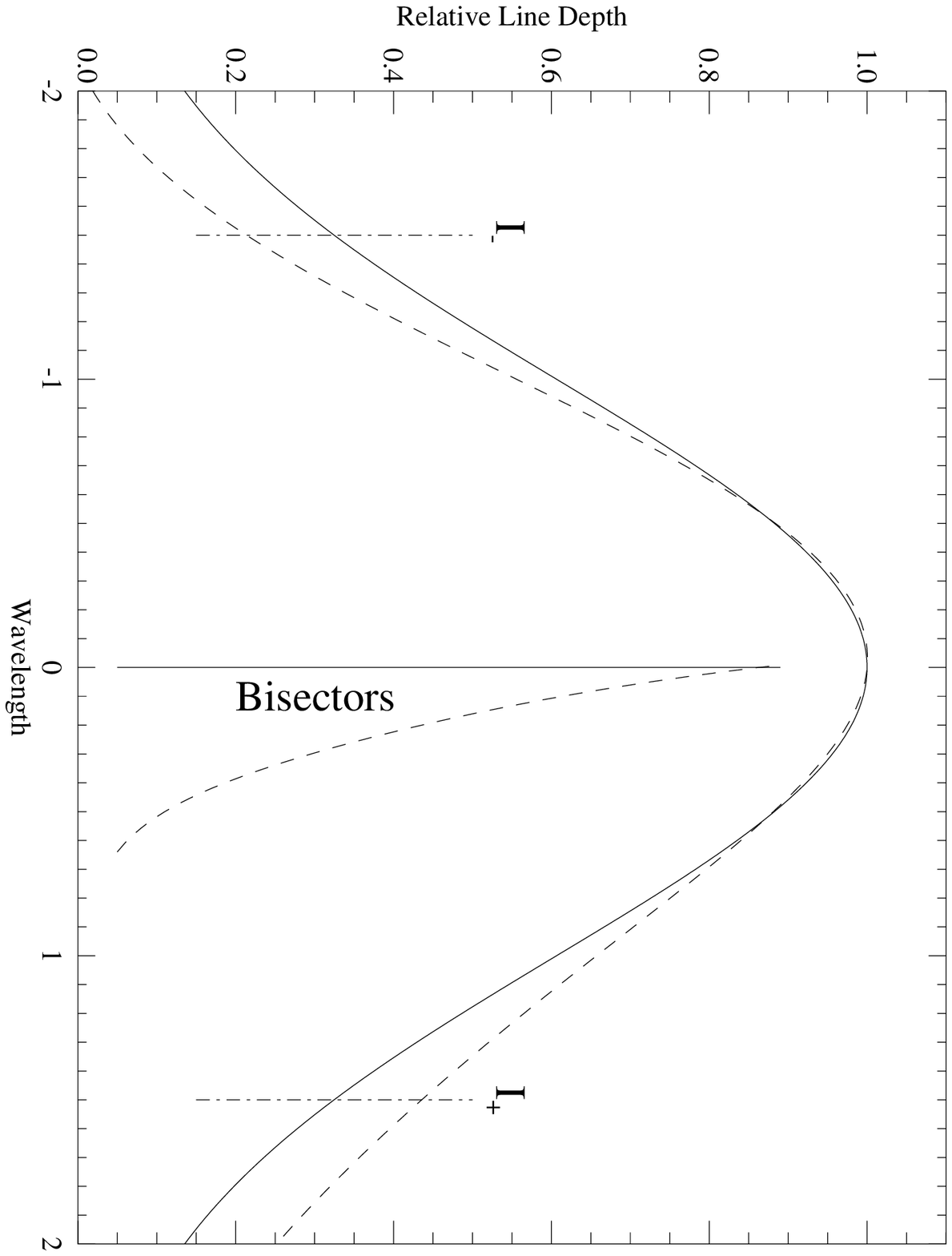,height=6.2in}
\ref Figure 1.  Geometry for computing line bisectors, and for
estimating the bisector shift resulting from a perturbation to
the line intensity.
An emission profile is shown, since the independent variable used
in the text ({\it i.e.,} relative line intensity) is most
conveniently represented in this way.
The solid curve is a Gaussian, whose vertical bisector is
shown as the solid line at zero wavelength.
The dashed curve is the same profile with the addition of
perturbations proportional to $\SH_1$ and $\SH_5$.
The corresponding bisector is the dashed line near the profile
center.
The intensities at $I_-$, $I_+$ illustrate that bisector motion
is associated with an antisymmetric perturbation of the line
intensity.
Note that, for clarity,  the intensity perturbations in this
example have been made much larger than those encountered in
practice.
\vfil\eject

\psfig{figure=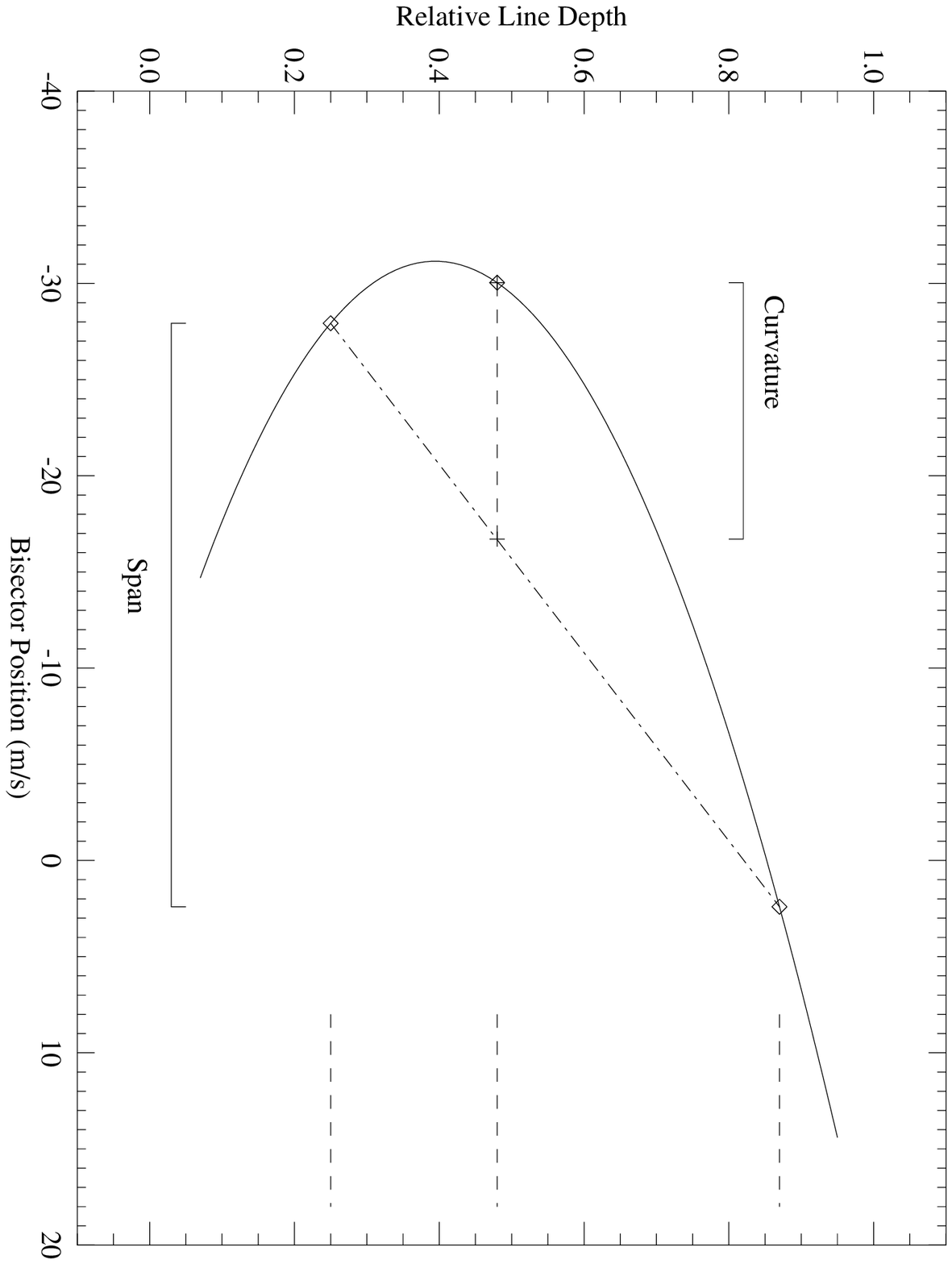,height=6.3in}
\ref Figure 2.  Definitions of the bisector span and curvature
used herein.
The bisector positions at relative line intensities of 0.25,
0.48, and 0.87 are indicated by diamond symbols.
The span is the horizontal distance between the top and
bottom samples;
the curvature is the difference between the center sample
and the linear interpolation between top and bottom samples,
evaluated at the relative intensity of the center point.
\vfil\eject

\psfig{figure=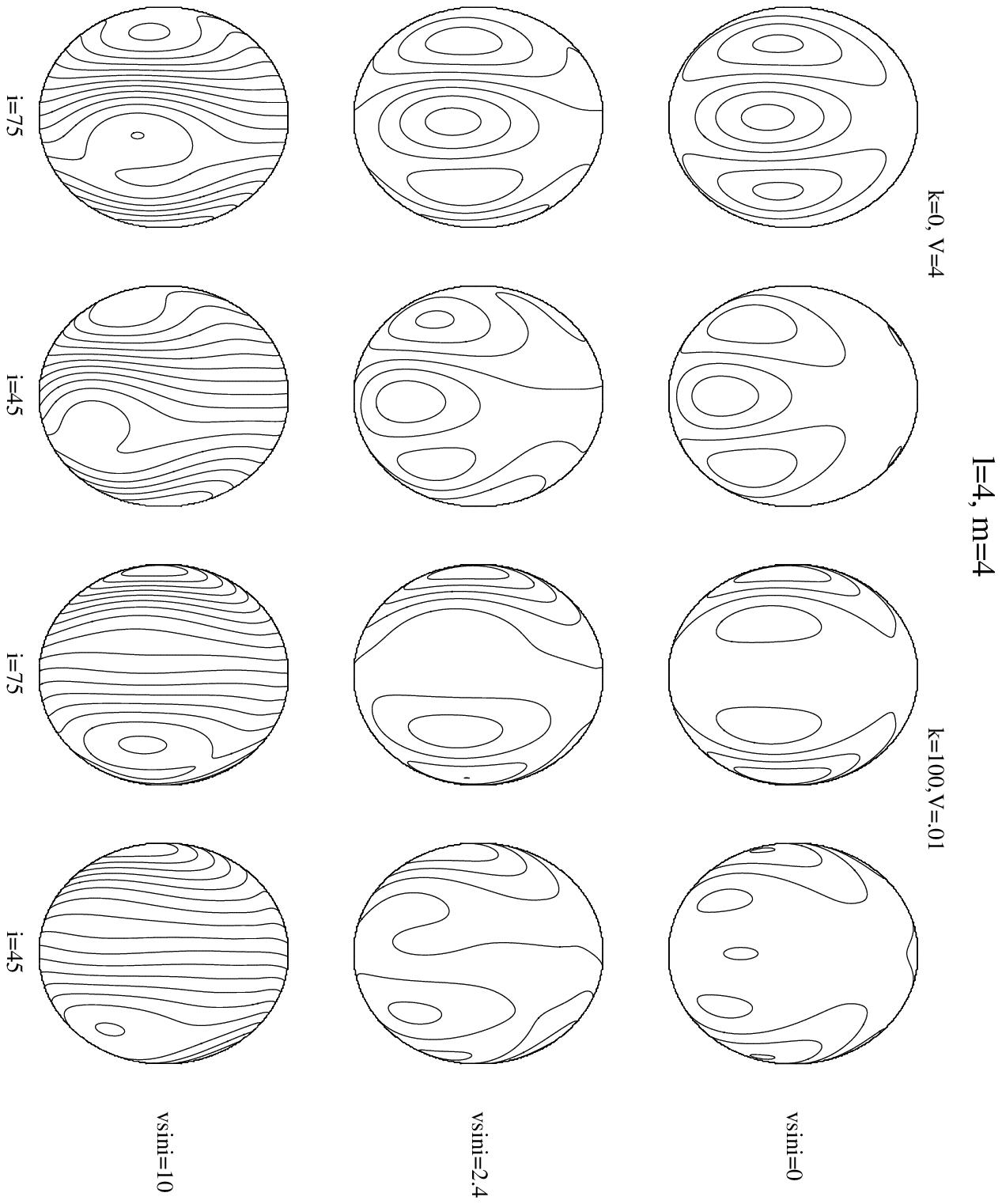,height=6.4in}
\ref Figure 3.  Contour maps of line-of-sight velocity for $\ell = 4$,
$m = 4$.
The three rows correspond to $v \sin i$ = 0, 2.4, and 10 km s$^{-1}$, from
top to bottom.
The leftmost two columns show $k=0$, $V_p$=4 km s$^{-1}$;
the rightmost two columns show $k=100$, $V_p$=10 m s$^{-1}$.
Peak surface flow speeds are the same in all cases.
Alternate columns correspond to $i=75 \degree$ and $i=45 \degree$.
The contour interval is 1.2 km s$^{-1}$.
For clarity of display, the velocity amplitudes chosen here are larger by
a factor of about 4 than any that have been suggested for 51 Peg.
In the absence of limb darkening, the Doppler shift density of
Eq. (21) would be proportional to the area between adjacent contours.
\vfil\eject

\psfig{figure=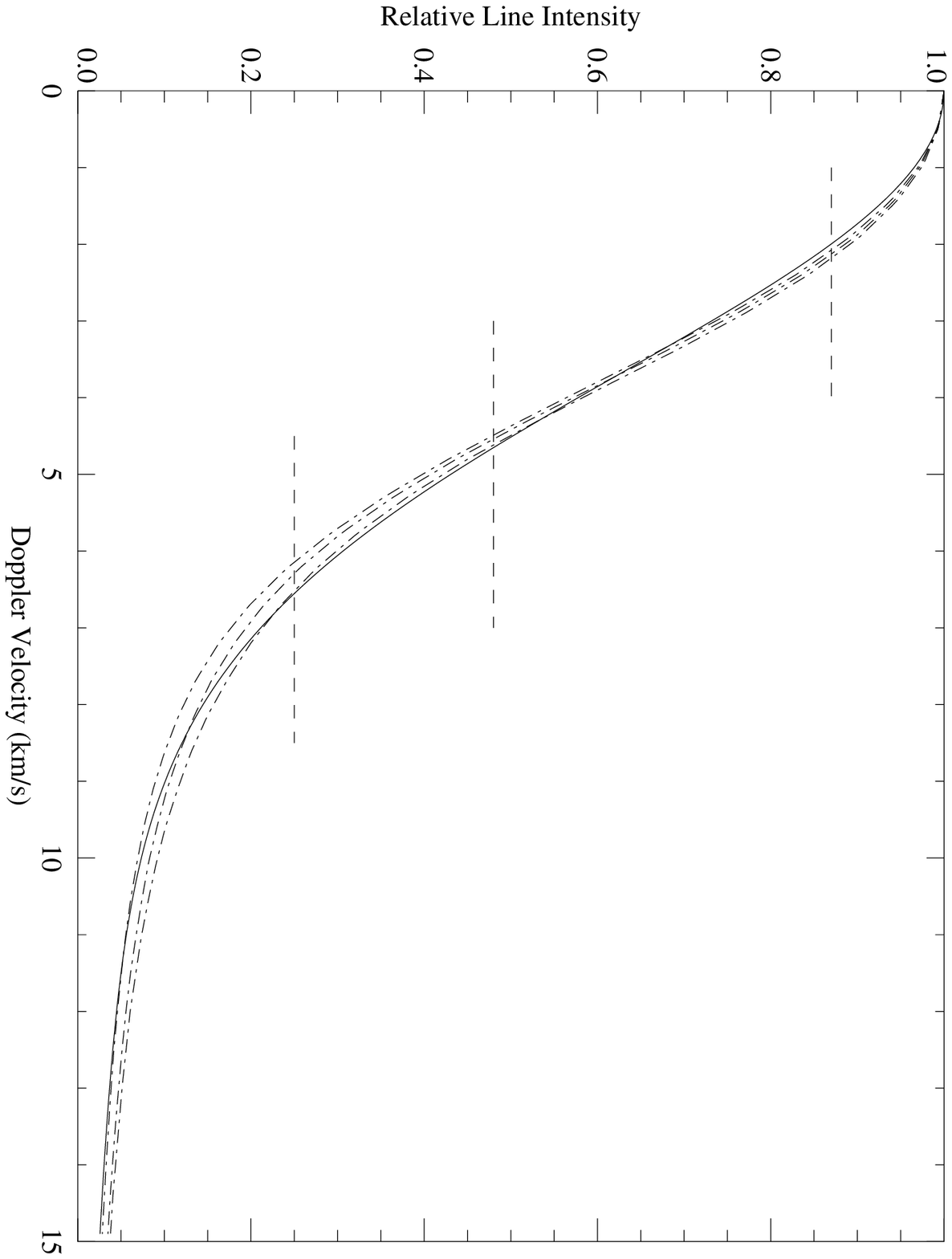,height=6.75in}
\ref Figure 4.  Half line profiles for two-Gaussian model 
with FWHM = 8.5 km s$^{-1}$ (solid line)
and for models used by Hatzes (1996)
for $\mu=\{$0, 0.5, 0.8 $\}$,
incorporating line transfer physics, including 3 km s$^{-1}$
microturbulence (dot-dashed lines).
All profiles have been scaled to the same central intensity,
so that their shapes may be compared.
Relative line intensities $U = \{$0.87, 0.48, 0.25$\}$,
which are used in the computation of bisector span and curvature,
are shown as horizontal dashed lines.
\vfil\eject

\psfig{figure=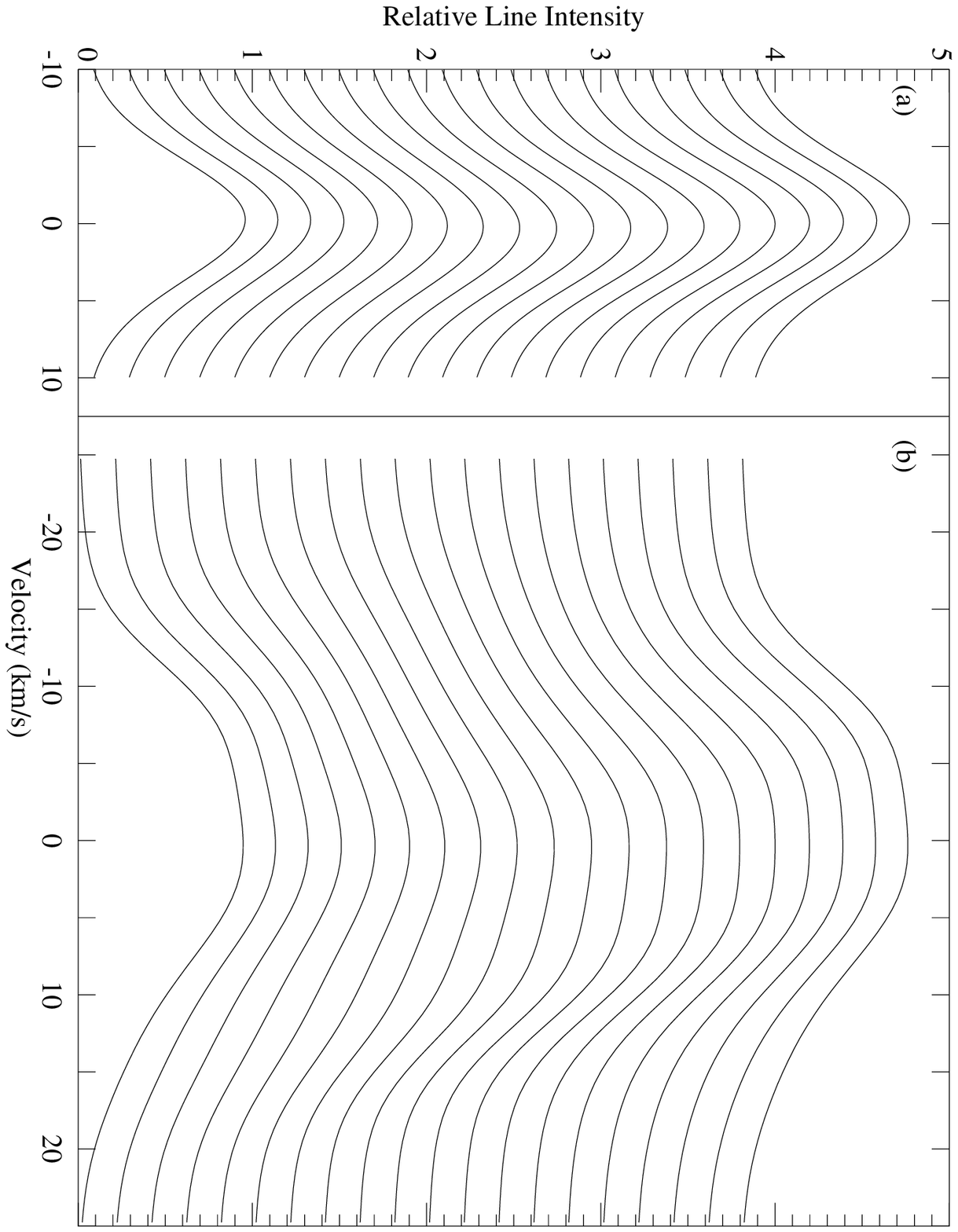,height=6.75in}
\ref Figure 5.  Time series of line profiles computed with 2-Gaussian
intrinsic profiles, $\ell = m = 4$, $i=75\degree$, $k=100$,
and $V_p=10$ m s$^{-1}$.
For the left panel, $v \sin i = 2.4$ km s$^{-1}$;
for the right, $v \sin i = 15$ km s$^{-1}$.
Time runs from bottom to top, through one oscillation cycle.
\vfil\eject

\psfig{figure=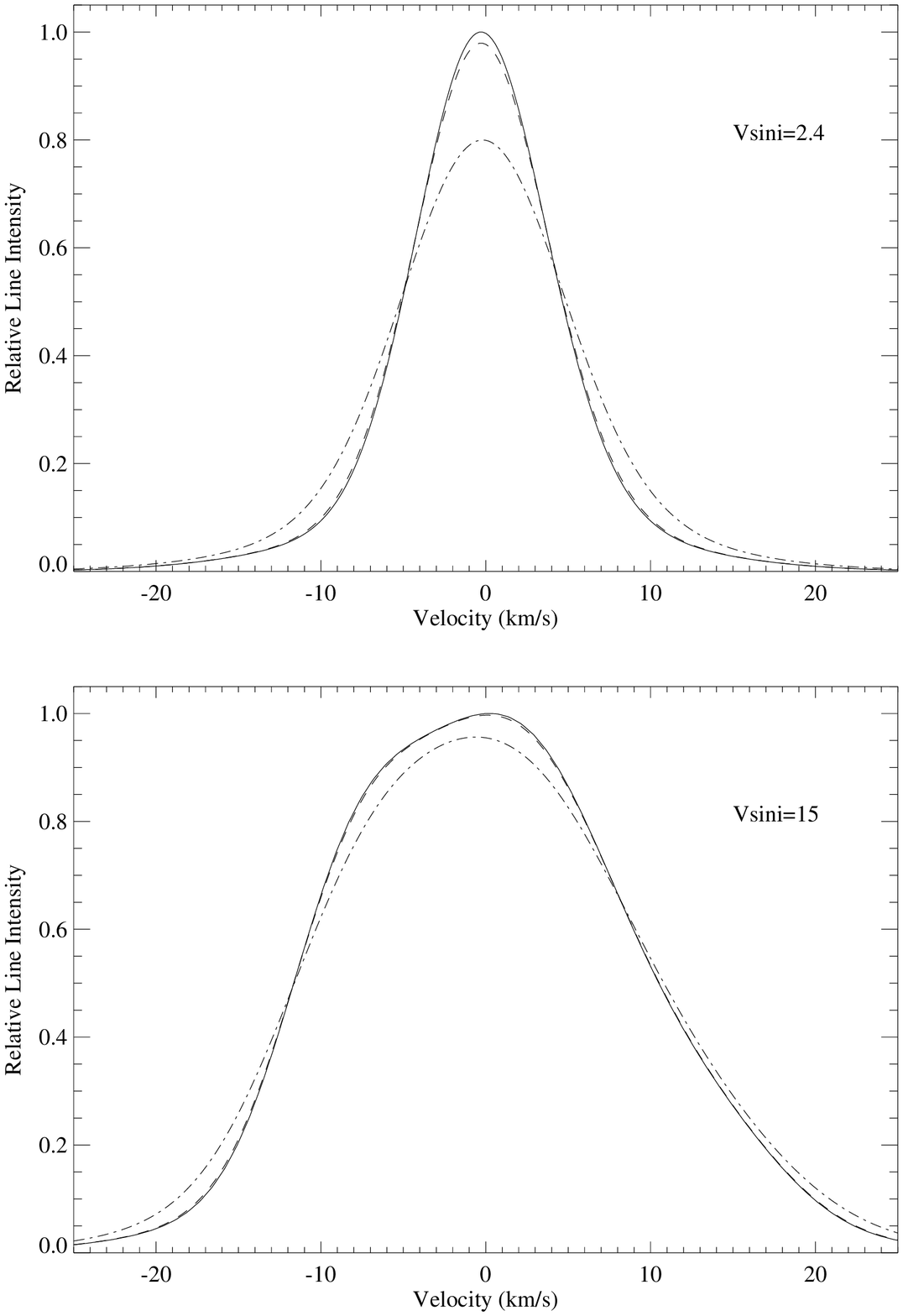,height=6.5in}
\ref Figure 6.  Line profiles broadened by stellar rotation
and by instrumental smearing.
Top: $v \sin i$=2.4 km s$^{-1}$;
bottom: $v \sin i$=15 km s$^{-1}$.
In both plots the solid profile is without instrumental smearing,
the dashed is smeared with a Gaussian PSF corresponding to
$R=10^5$,
and the dot-dashed profile is smeared with the measured AFOE
instrumental PSF, corresponding approximately to $R=50000$.
These profiles have been scaled to unit central intensity
for the case without instrumental smearing.
\vfil\eject

\psfig{figure=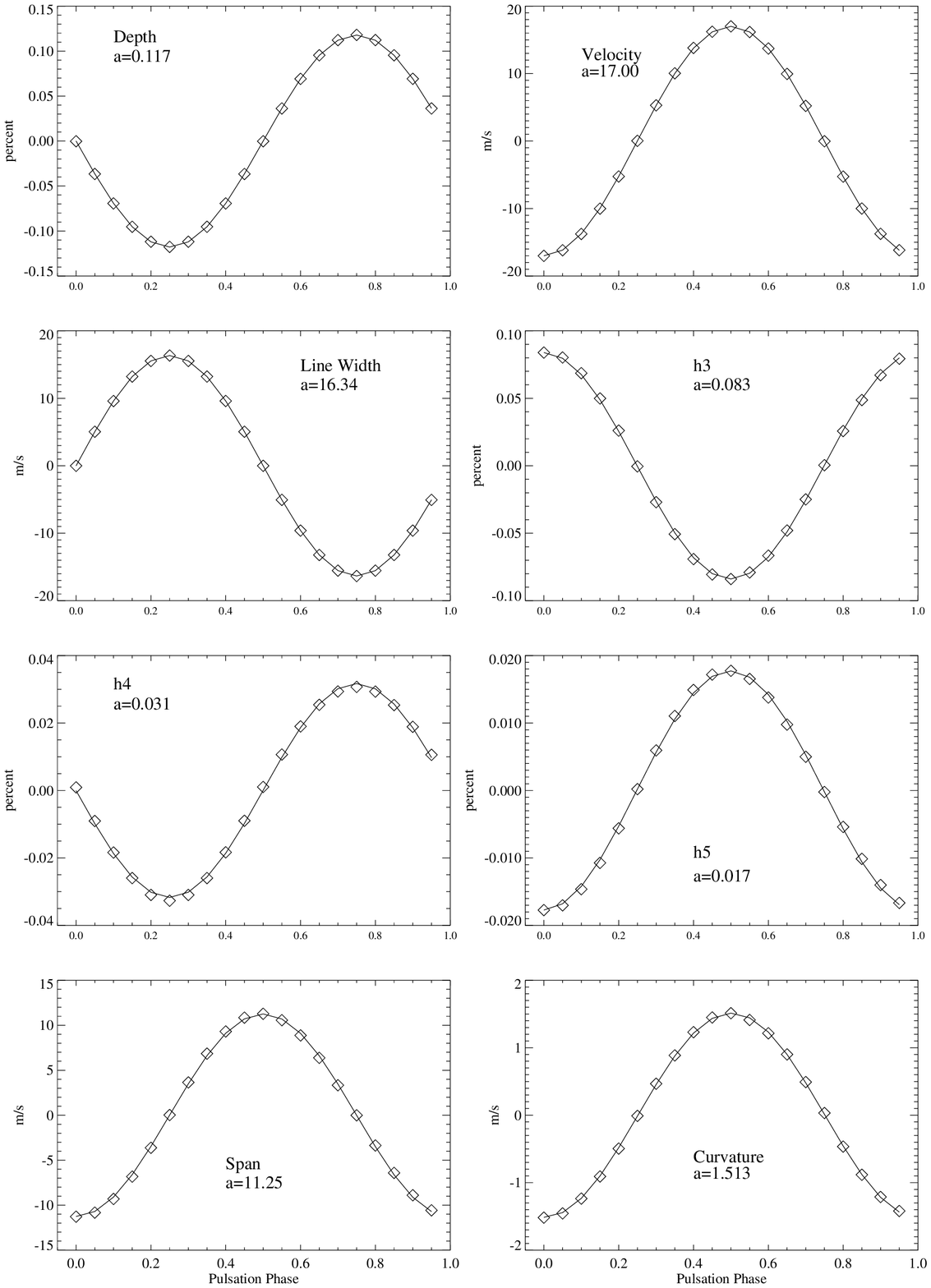,height=6.75in}
\ref
Figure 7.  Time series of the 8 parameters describing simulated line shape
and position during one cycle of pulsation.
Parameters for this simulation were $\ell=m=4$, $V_p$=1 m s$^{-1}$,
$k=100$, $i=75\degree$, $v \sin i=2.4$ km s$^{-1}$.
Note that $V_p$ is 10 times smaller than in the previous figures,
to make the RV signal more nearly match observed values.
Calculated points are shown as diamonds, with sinusoidal fits
shown as solid lines.
The fitted amplitudes (in percent or in m s$^{-1}$) are displayed
in each panel.
\vfil\eject

\psfig{figure=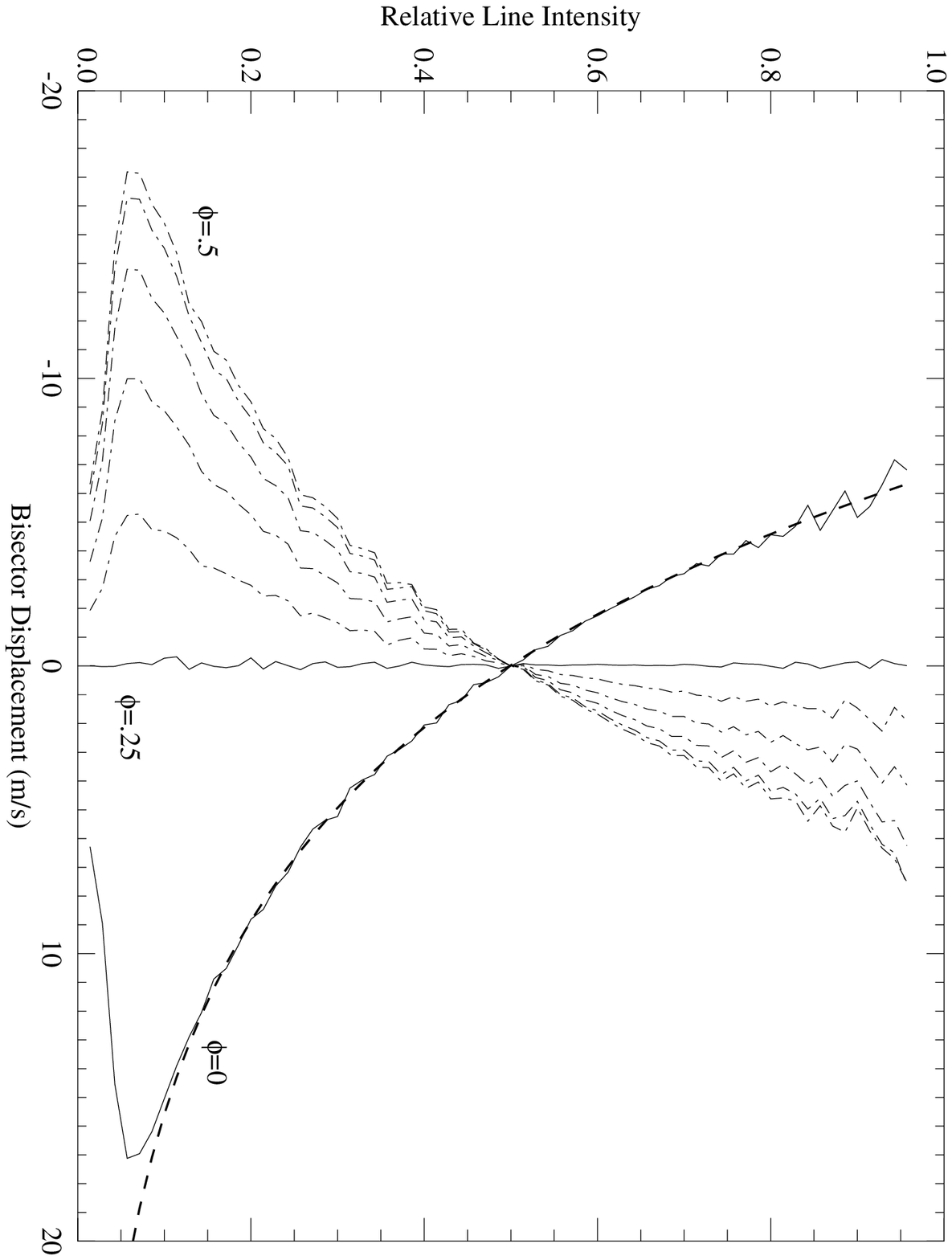,height=6.5in}
\ref
Figure 8.  Simulated line bisector positions $\delta B$ 
for 2-Guassian models and a range of oscillation phases,
spanning half an oscillation cycle,
shown as a function of relative line intensity $U$.
For this simulation $V_p=1$ m s$^{-1}$, $\ell=m=4$,
$i=75\degree$.
The phases shown are $\phi= \{$0., .25, .3, .35, .4, .45, .5$\}$.
Also shown, for $\phi=0$, is a fit to a function of the form
$\delta B \ = \ A \ + B \ln U$ (heavy dashed line),
weighted so that points with $U \leq 0.07$
are excluded from the fit.
The small magnitude of the $\delta B$ values at $U \leq 0.06$ result from the
strong wings of the 2-Gaussian profiles.
\vfil\eject

\psfig{figure=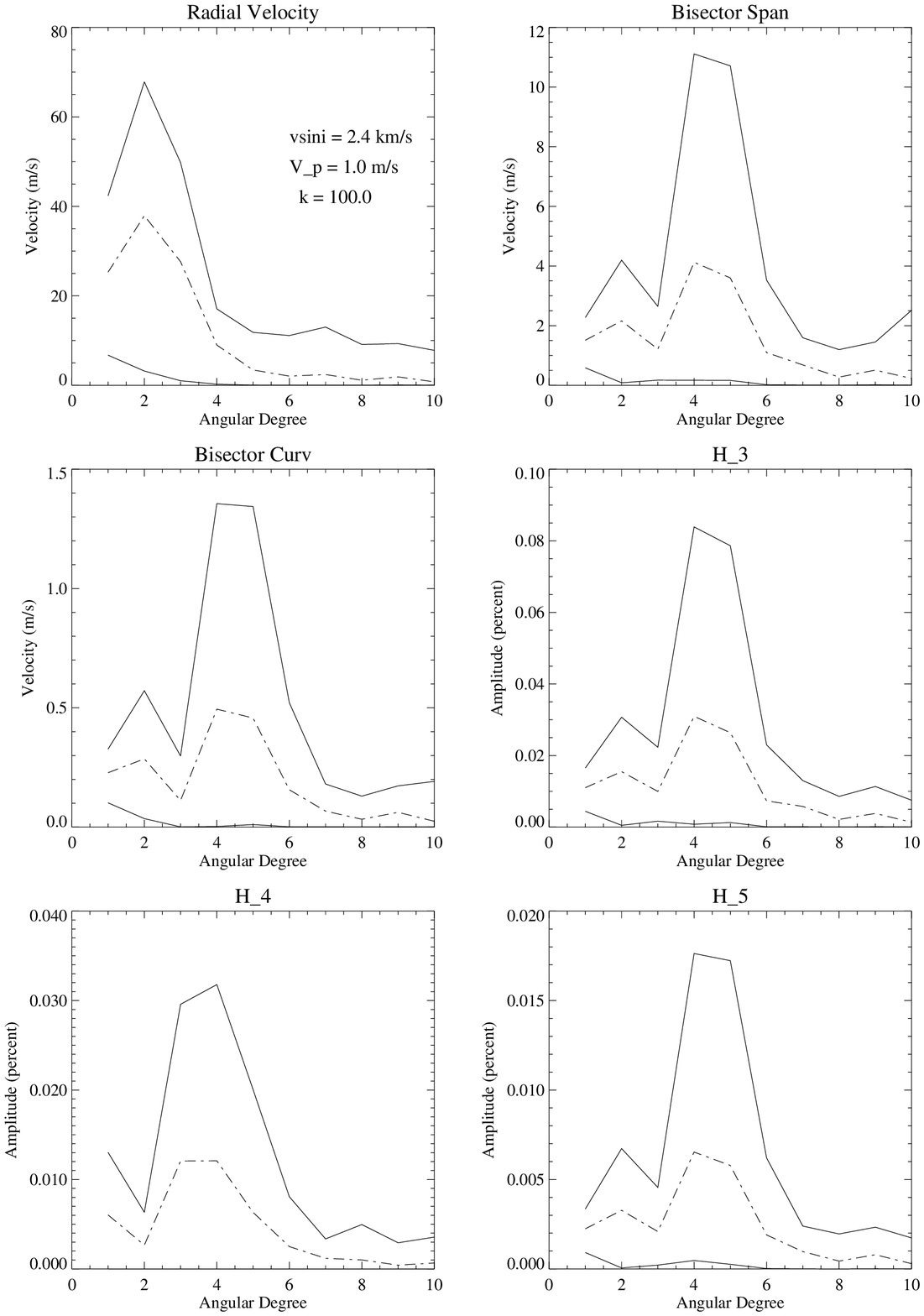,height=6.2in}
\ref
Figure 9. Amplitudes of the sinusoidal variation of six of the 
parameters describing simulated
line shape and position, shown as a function of $\ell$.
The remaining 2 parameters (line depth and width) have amplitudes
that are too small to measure conveniently, and contain no
additional information.
3 values are plotted at each $\ell$, namely the maximum, minimum, and
mean of the amplitude for the given parameter, taken over the
range of $m$ and $i$ values for which simulations were computed.
In all cases the mean values are indicated with dot-dashed lines.
This simulation used parameters $v \sin i = 2.4$ km s$^{-1}$,
$V_p=1$ m s$^{-1}$, $k=100$.
\vfil\eject

\psfig{figure=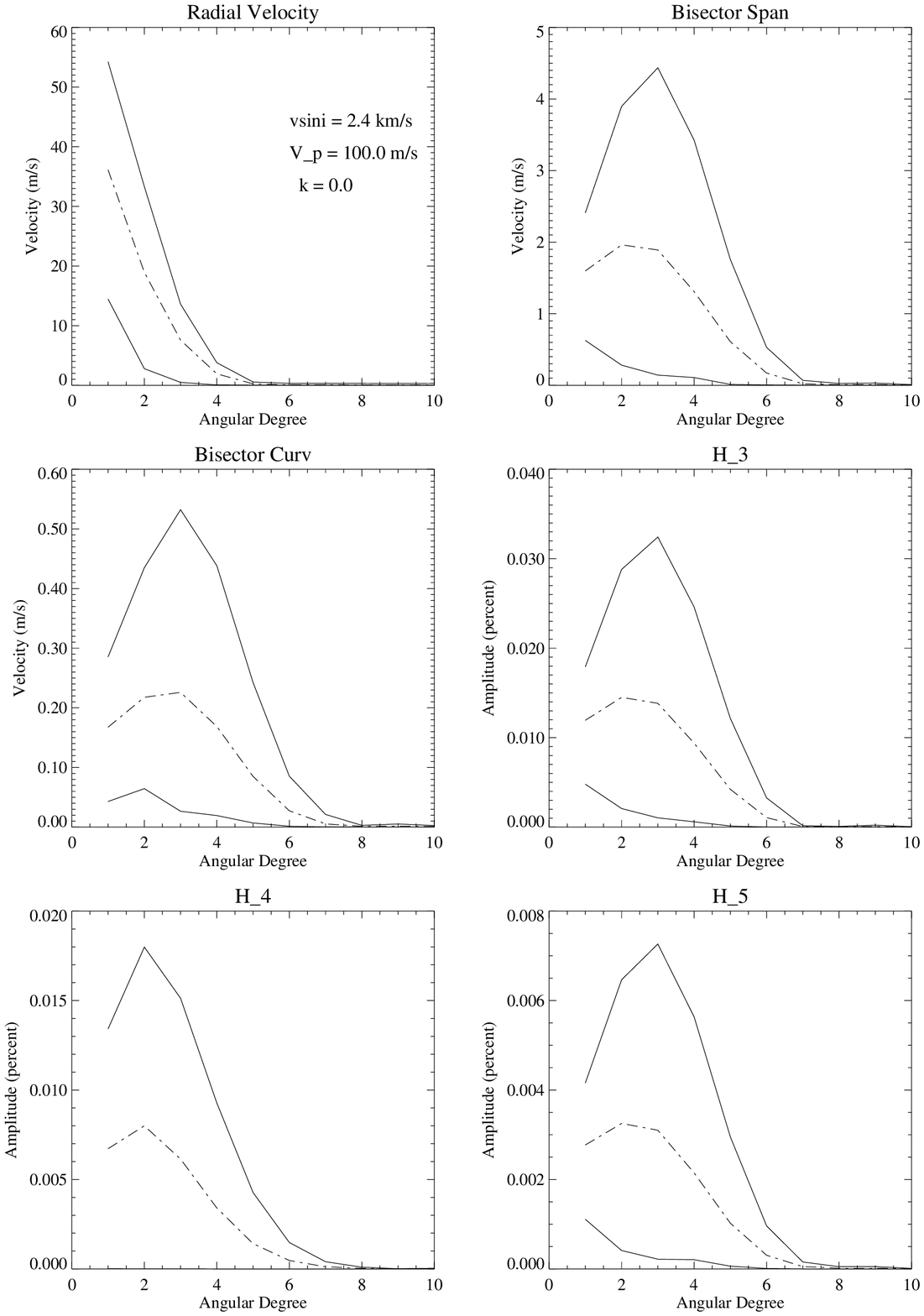,height=7.5in}
\ref
Figure 10.  Same as Fig. 9, except $v \sin i = 2.4$ km s$^{-1}$,
$V_p = 100$ m s$^{-1}$, $k=0$.
\vfil\eject

\psfig{figure=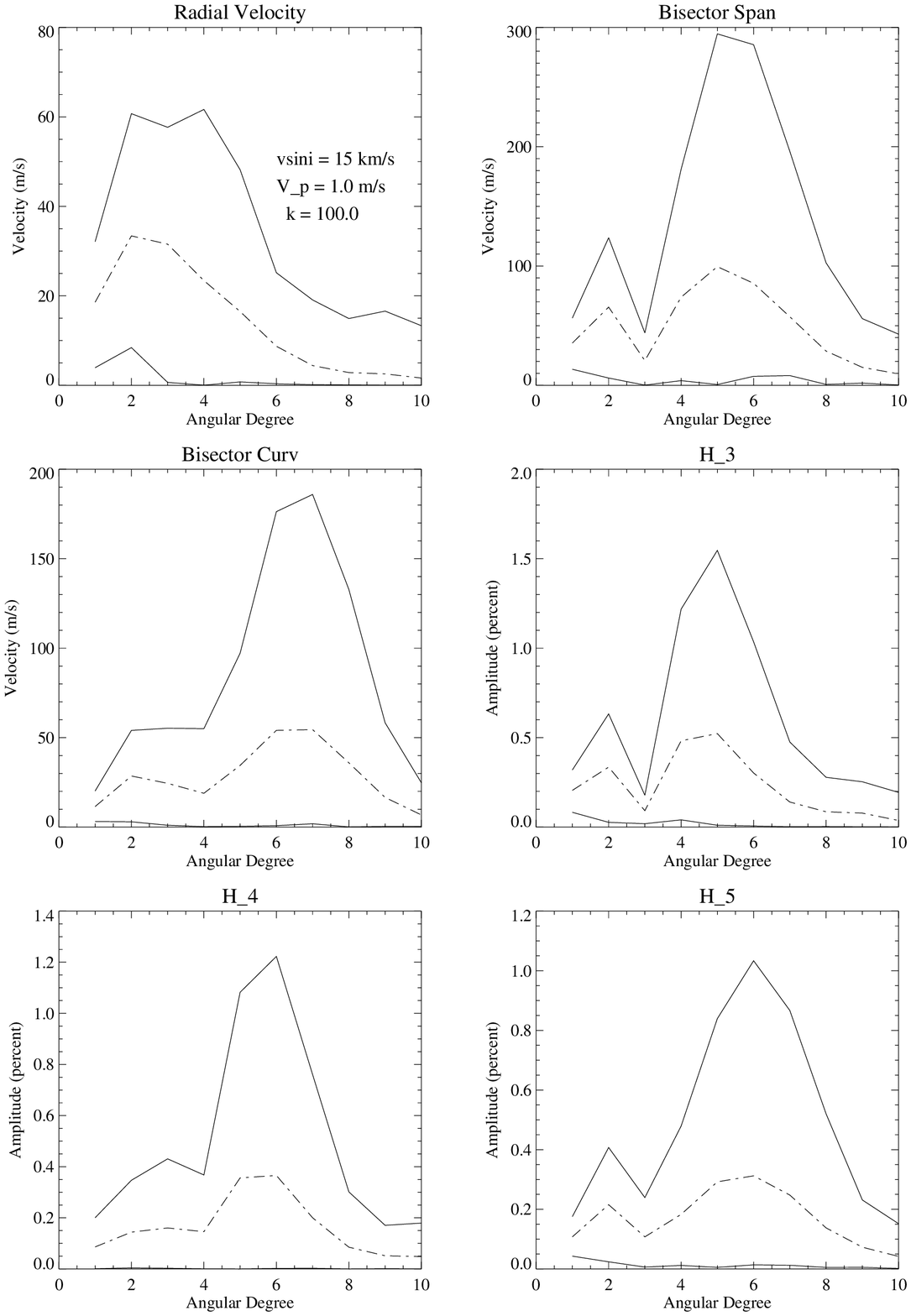,height=7.5in}
\ref
Figure 11.  Same as Fig. 9, except $v \sin i = 15$ km s$^{-1}$,
$V_p = 1$ m s$^{-1}$, $k=100$.
\vfil\eject

\psfig{figure=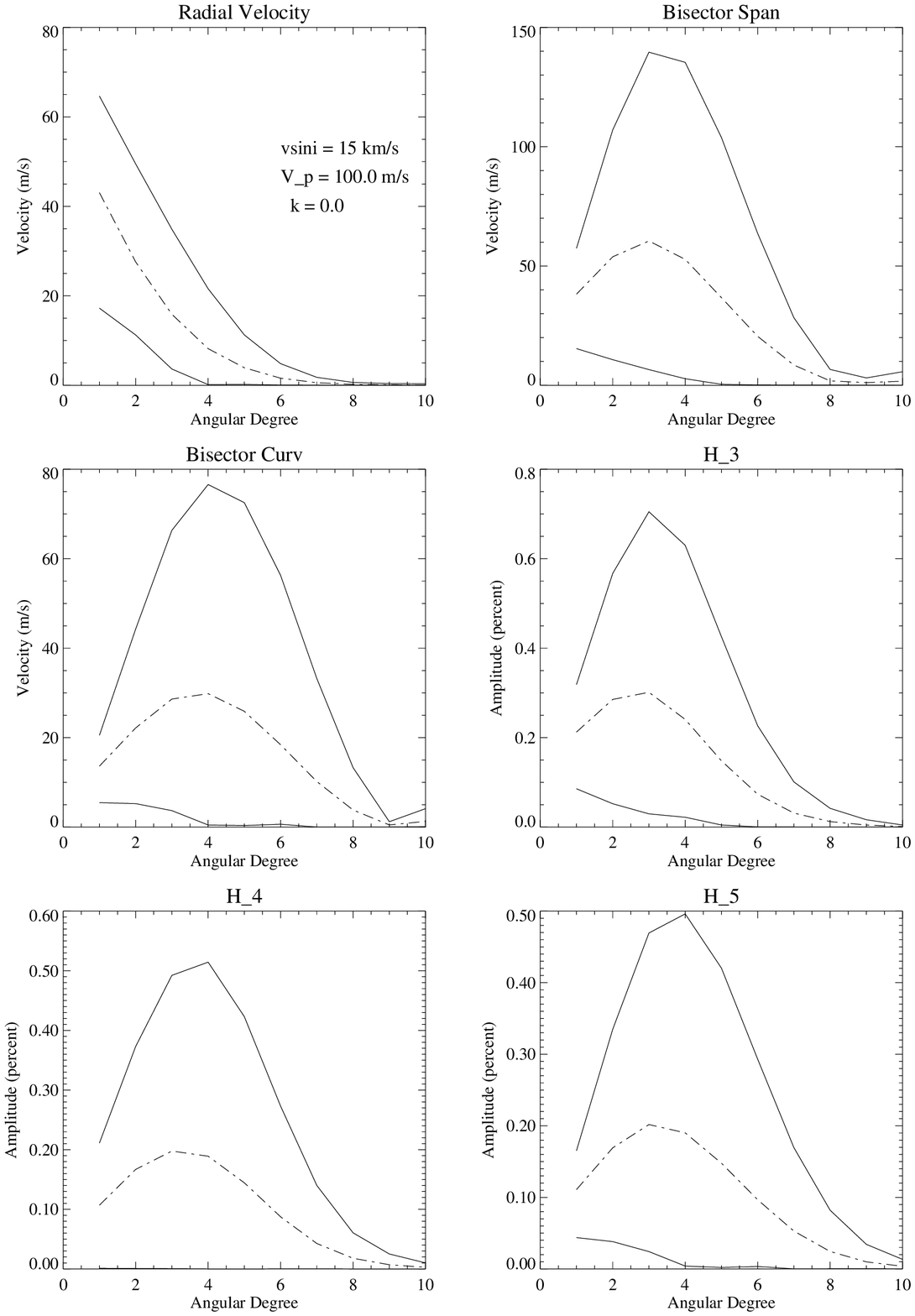,height=7.5in}
\ref
Figure 12.  Same as Fig. 9, except $v \sin i = 15$ km s$^{-1}$,
$V_p = 100$ m s$^{-1}$, $k=0$.
\vfil\eject

\psfig{figure=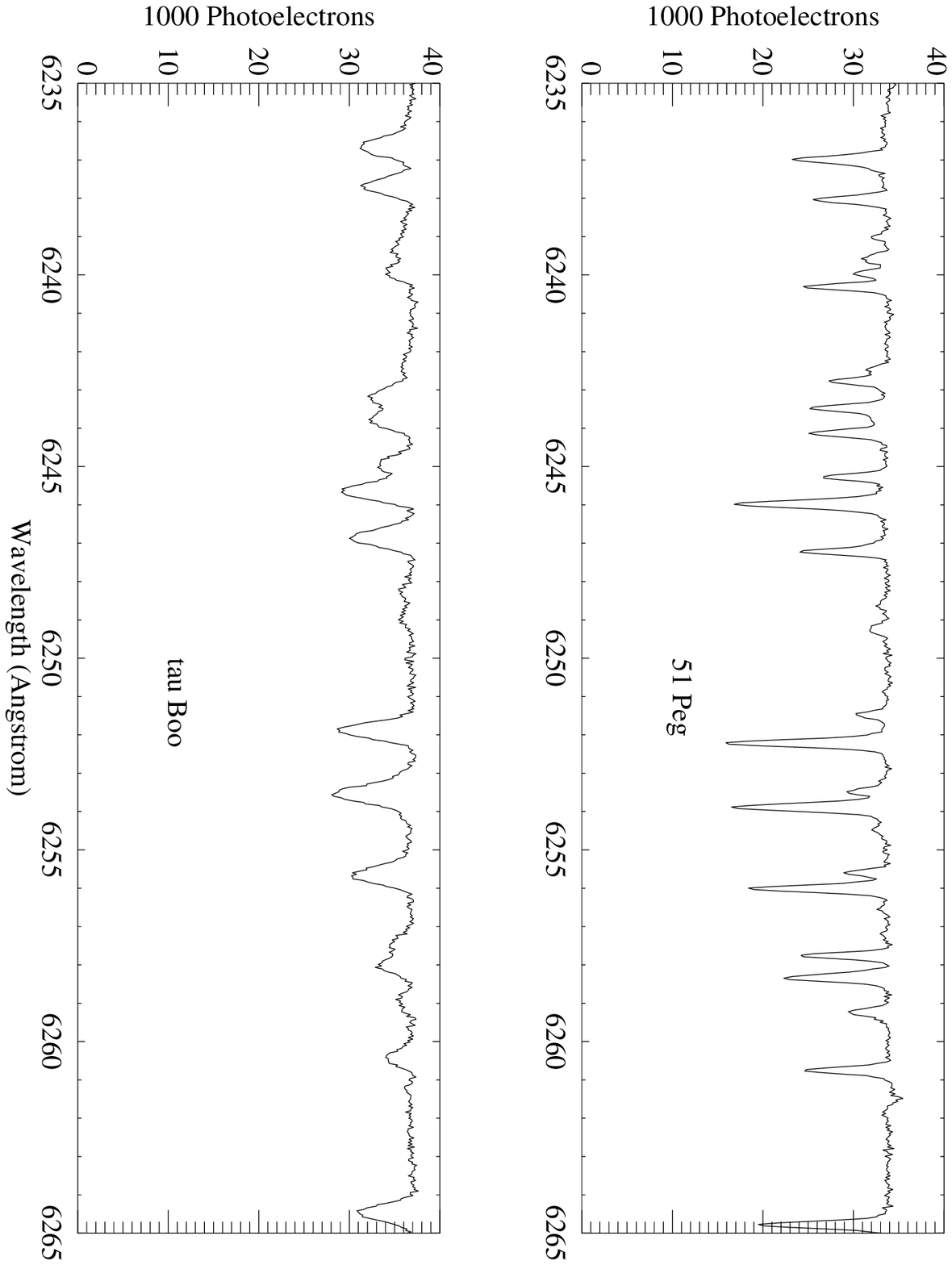,height=7.0in}
\ref
Figure 13.  Representative segments of observed spectra for
51 Peg (top) and $\tau$ Boo (bottom), for 600 s integration times.  
These spectra have been corrected for detector bias and gain
variations, and are displayed in units of detected photoelectrons.
\vfil\eject

\psfig{figure=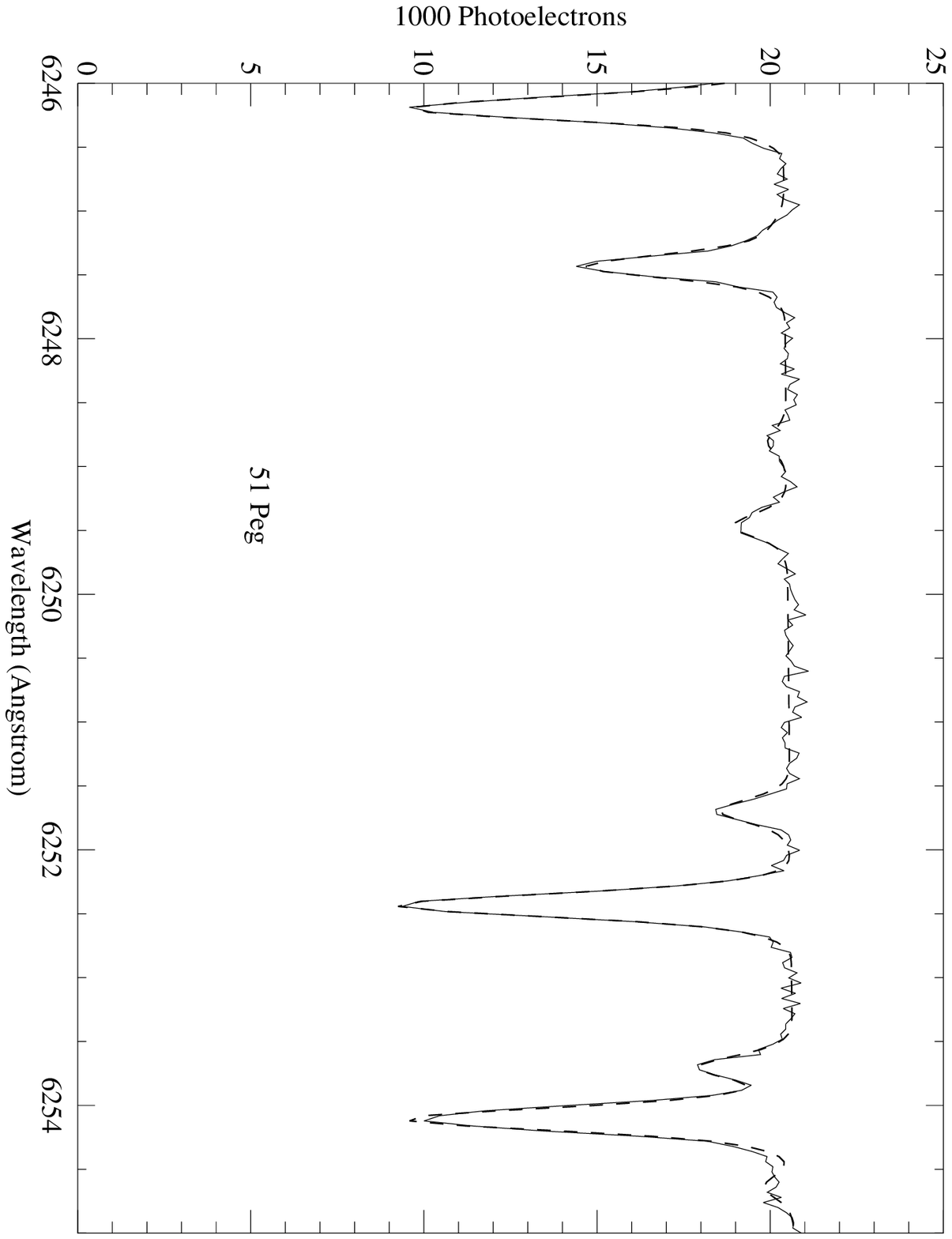,height=7.2in}
\ref
Figure 14.  Expanded plot of a portion of a spectrum of 51 Peg
in the region near the $\lambda$6253 Fe I line.
Overplotted (heavy line) is the multi-Gaussian model fit
to these data.
\vfil\eject

\psfig{figure=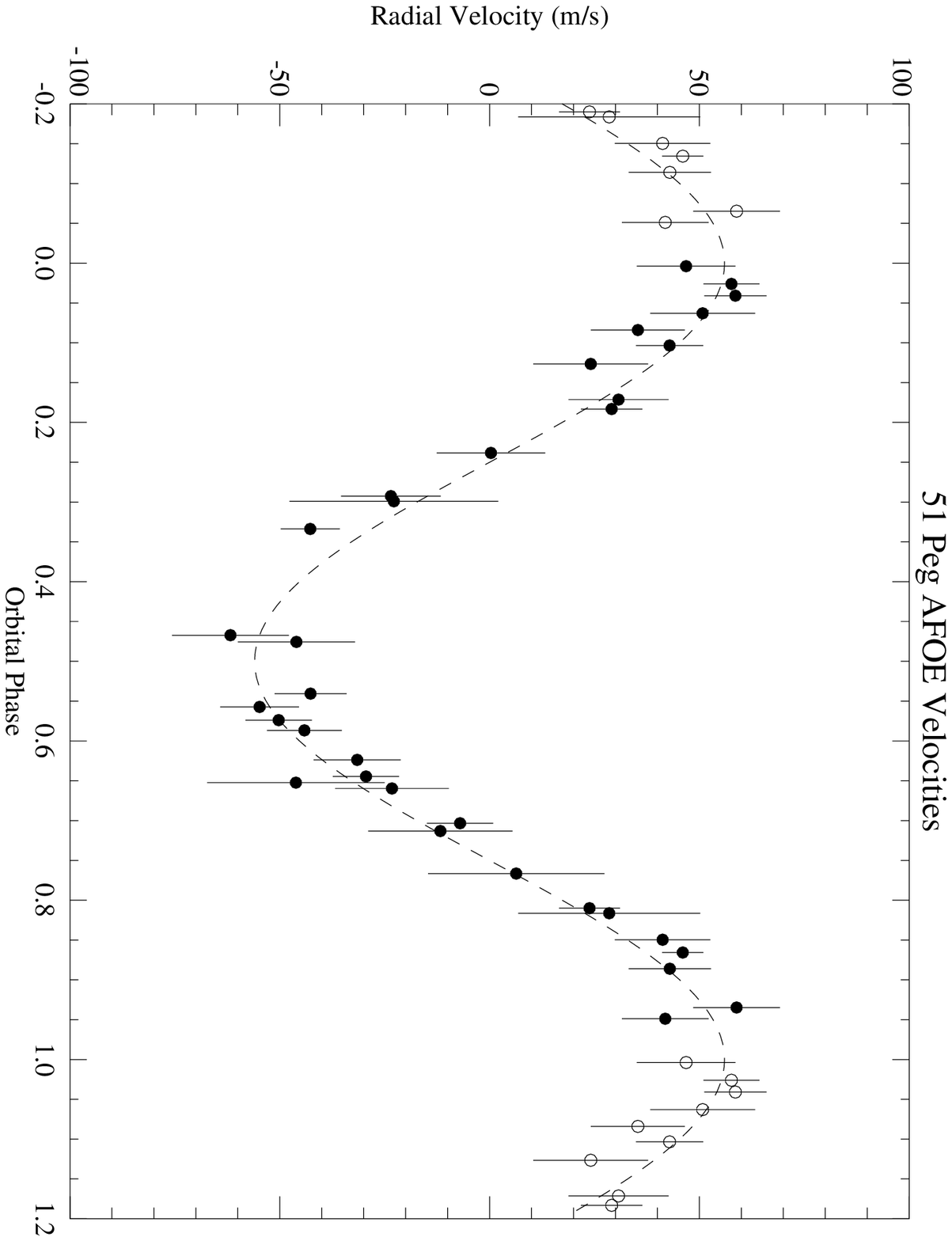,height=6.4in}
\ref
Figure 15.  The radial velocity of 51 Peg as measured by the AFOE
between July 1995 and June 1997.
plotted against orbital phase calculated from the 
ephemeris of Mayor \& Queloz (1996).
Ephemeris predictions are plotted as the solid line.
The filled circles span one complete RV cycle, hence encompass
all of the independent data points.
The open circles are redundant and cover an additional span
in phase, so that the phase of the observed sinusoid may be more
easily visualized.
Month-to-month offsets have been applied to the earlier data
because of long-term instrumental drifts (now largely eliminated);
however these drift corrections have negligible effect on the
measurement of RV variations with a 4.2 day period.
\vfil\eject

\psfig{figure=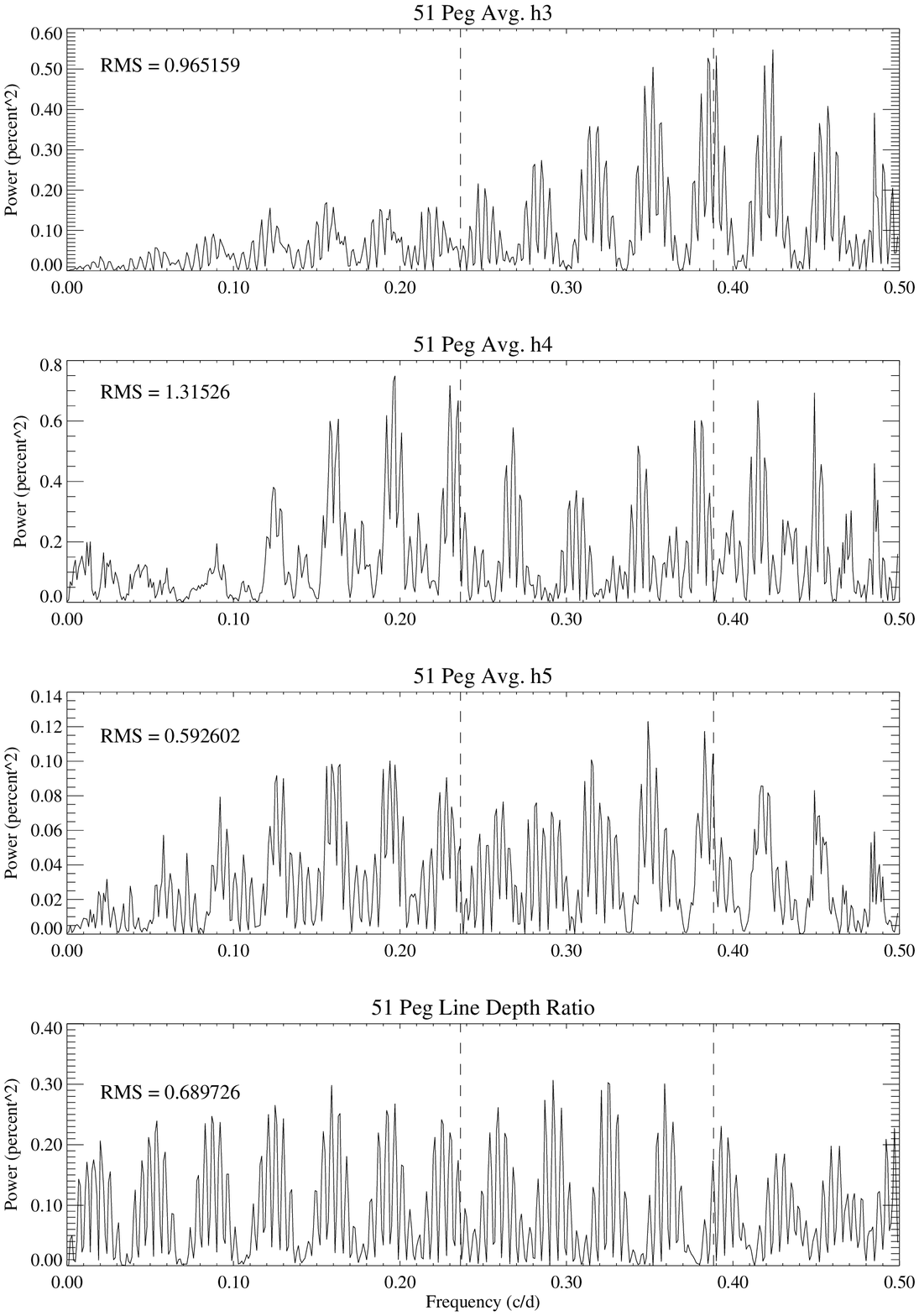,height=6.75in}
\ref
Figure 16.  Same as Fig. 15, except that the plotted quantities are
(a) line shape coefficient $h_3$;
(b) line shape coefficient $h_4$;
(c) line shape coefficient $h_5$;
(d) line depth ratio $\lambda$6252 V I to $\lambda$6253 Fe I.
In this figure, the solid curves are the sinusoids resulting
from unweighted least-squares fits to the time series,
assuming a period equal to the RV period.
The amplitudes of these sinusoids (in percent) are displayed on the plots.  
\vfil\eject

\psfig{figure=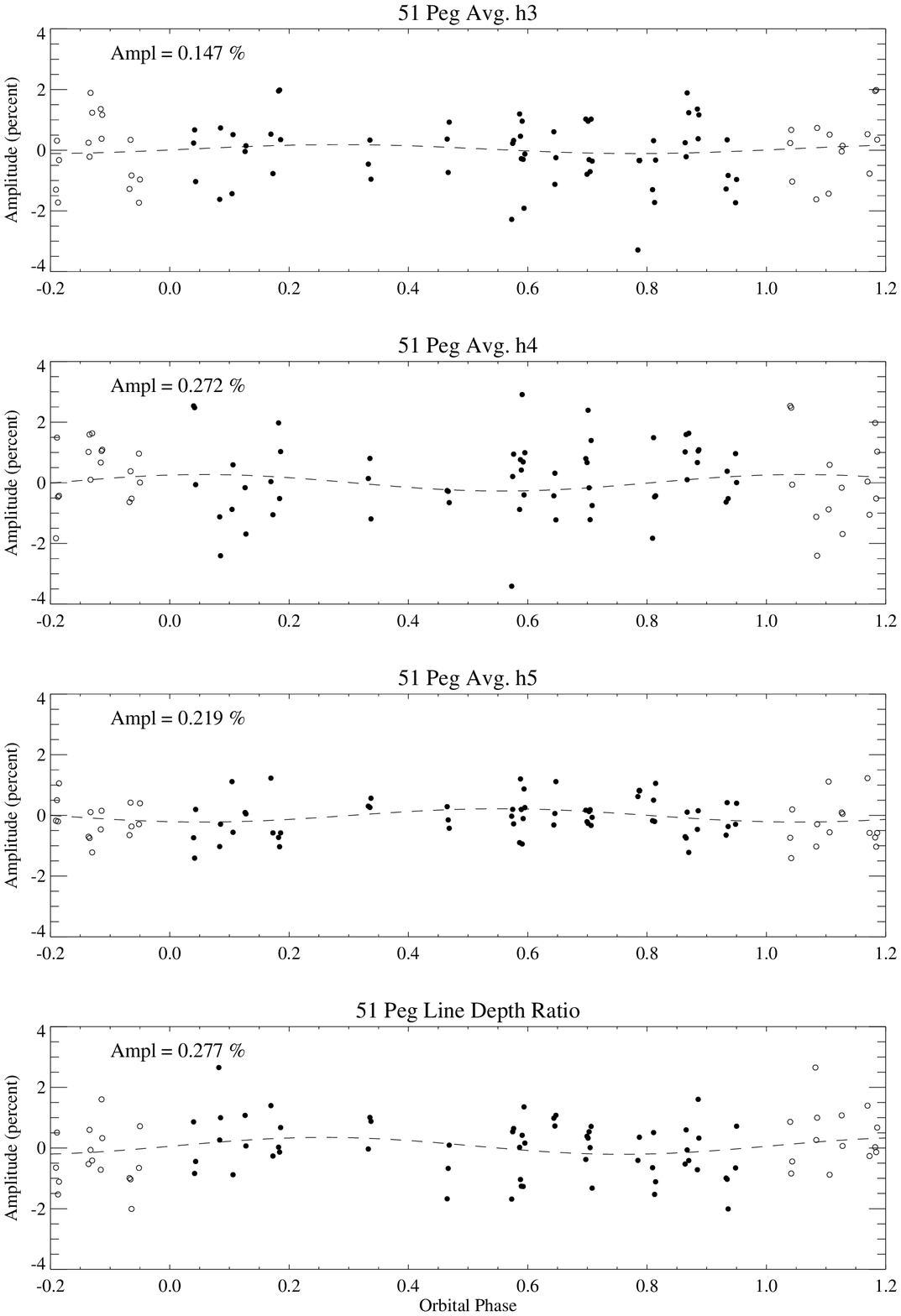,height=6.75in}
\ref
Figure 17.  Periodograms of the data displayed in Fig. 16.
The spectra are normalized so that a signal with an amplitude of 1\%
would appear as a power concentration whose central peak has unit height.
Frequencies are shown in cycles per day.
Vertical dashed lines indicate the frequencies reported by GH,
corresponding to periods of 4.231 and 2.575 days.
\vfil\eject

\psfig{figure=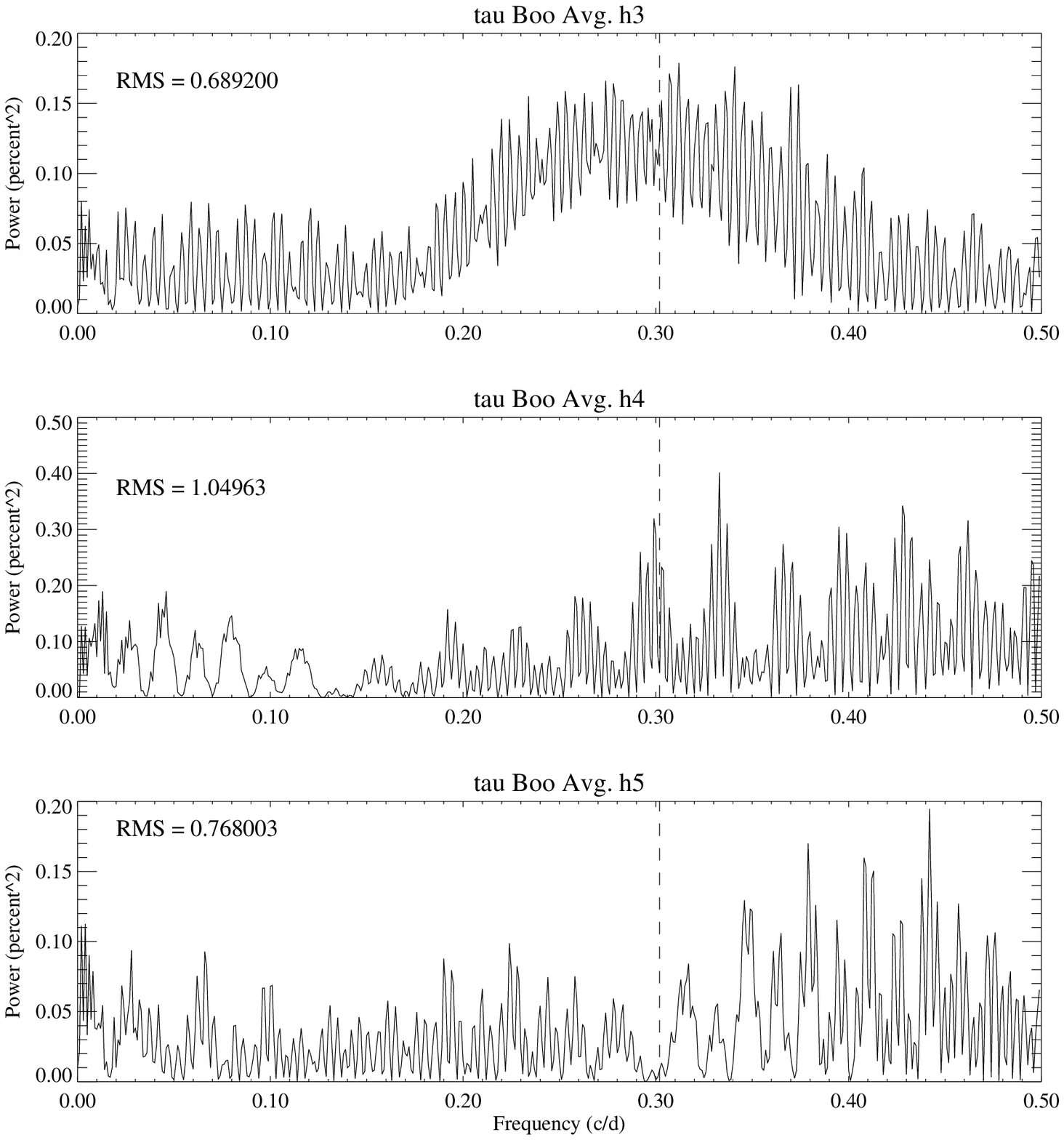,height=7.5in}
\ref
Figure 18.  Same as Fig. 16, except the observations are of $\tau$ Boo.
Also, the line depth ratio is not plotted, since this ratio is not
useful for $\tau$ Boo.
\vfil\eject

\psfig{figure=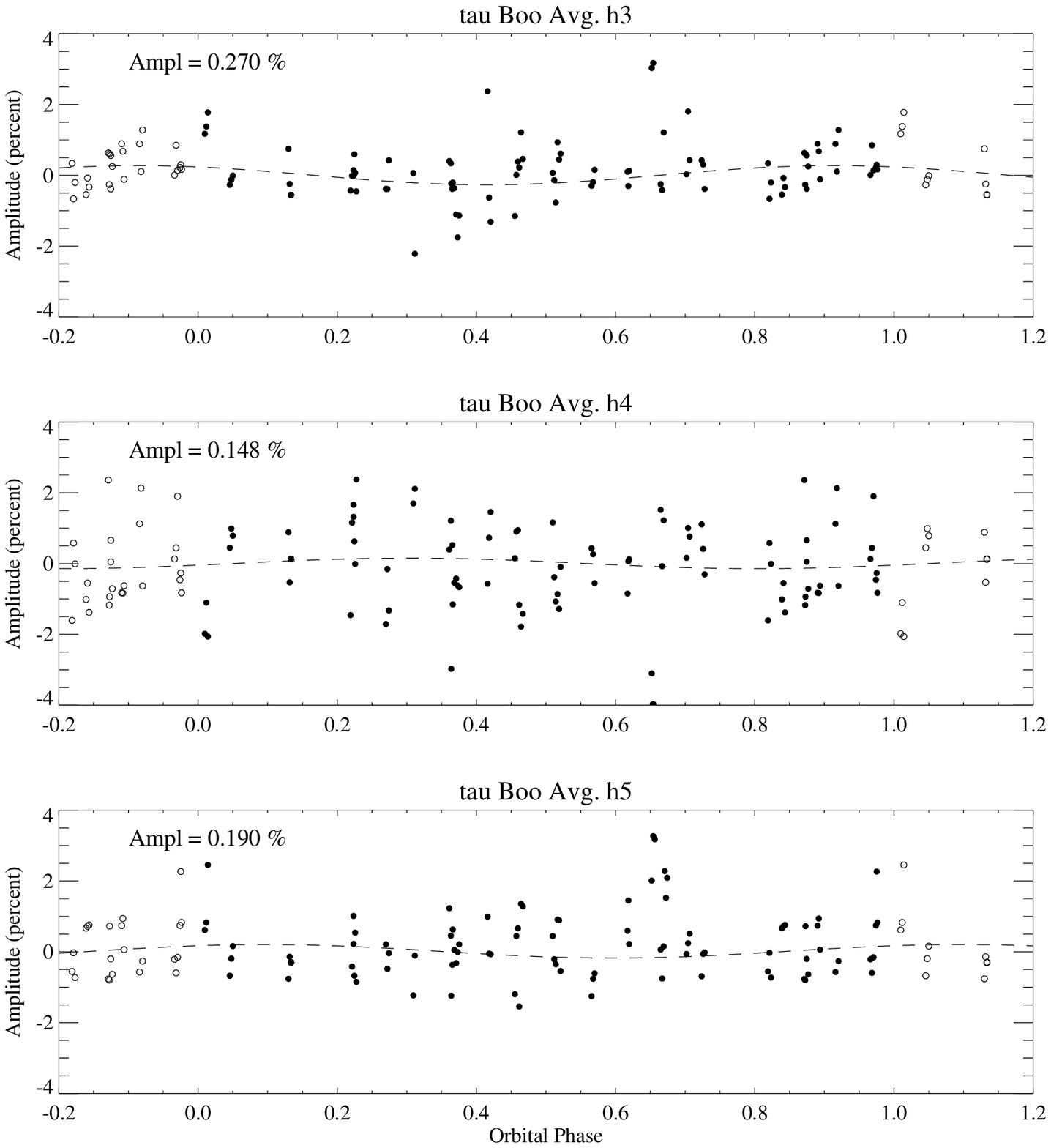,height=7.5in}
\ref
Figure 19.  Same as Fig. 17, except the observations are of $\tau$ Boo.
As in Fig. 18, line depth ratio data are not plotted.
\vfil\eject

\bye